\documentclass[amsmath,amssymb,floatfix,aps,pra,twocolumn,superscriptaddress]{revtex4-2}

\usepackage{xurl}
\usepackage{graphicx}%
\usepackage{dcolumn}%
\usepackage{layouts}
\usepackage{floatrow}
\usepackage{amsmath}
\DeclareMathOperator{\Tr}{Tr}

\usepackage[caption=false,font=footnotesize]{subfig}
\usepackage{braket}

\usepackage{bm} %

\newcommand{\bx}{\mathbf{x}}

\begin{document}

\title{Numerical evidence against advantage with quantum fidelity kernels on classical data} 

\author{Lucas Slattery}
\affiliation{Department of Physics, University of Illinois Urbana-Champaign, Champaign, IL 61820 USA}
\affiliation{Mathematics and Computer Science Division, Argonne National Laboratory, Lemont, IL 60439 USA}
\author{Ruslan Shaydulin}
\affiliation{Global Technology Applied Research, JPMorgan Chase, New York, NY 10017 USA}
\author{Shouvanik Chakrabarti}
\affiliation{Global Technology Applied Research, JPMorgan Chase, New York, NY 10017 USA}
\author{Marco Pistoia}
\affiliation{Global Technology Applied Research, JPMorgan Chase, New York, NY 10017 USA}
\author{Sami Khairy}
\thanks{Now at Microsoft.}
\affiliation{Mathematics and Computer Science Division, Argonne National Laboratory, Lemont, IL 60439 USA}
\author{Stefan M.\ Wild}
\affiliation{Mathematics and Computer Science Division, Argonne National Laboratory, Lemont, IL 60439 USA}

\date{\today}

\begin{abstract}

Quantum machine learning techniques are commonly considered one of the most promising candidates for demonstrating practical quantum advantage. In particular, quantum kernel methods have been demonstrated to be able to learn certain classically intractable functions efficiently if the kernel is well-aligned with the target function. In the more general case, quantum kernels are known to suffer from exponential ``flattening'' of the spectrum as the number of qubits grows, preventing generalization and necessitating the control of the inductive bias by hyperparameters. We show that the general-purpose hyperparameter tuning techniques proposed to improve the generalization of quantum kernels lead to the kernel becoming well-approximated by a classical kernel, removing the possibility of quantum advantage. We provide extensive numerical evidence for this phenomenon utilizing multiple previously studied quantum feature maps and both synthetic and real data. Our results show that unless novel techniques are developed to control the inductive bias of quantum kernels, they are unlikely to provide a quantum advantage on classical data.
\end{abstract}

\maketitle

\section{\label{sec:intro}Introduction}

Quantum computers are known to have the theoretical potential to solve certain problems faster than their classical counterparts~\cite{nielsen2011quantum}. Machine learning problems are of particular interest due to their wide applicability and the possibility of large quantum speedups~\cite{Biamonte2017, Pistoia2021,2201.02773}.
Quantum machine learning methods applied to quantum data (i.e., data given as quantum states) are known to provide an exponential advantage over classical machine learning methods \cite{Huang2022,Kubler2021}. On the other hand, if the input data is classical (such as vectors of finite-precision numbers), exponential end-to-end quantum speedup for machine learning is only known for very specifically crafted data~\cite{Liu2021,Lloyd2022TDA}. The lack of provable guarantees for classical data motivates the development of techniques that improve the performance of quantum machine learning methods with the goal of empirically demonstrating quantum advantage.

A particular class of quantum machine learning methods we study in this work is quantum fidelity kernel methods applied to classical data. In quantum fidelity kernel methods, a classical data point is embedded in a quantum state using a parameterized quantum evolution (typically, a parameterized quantum circuit) called a quantum feature map~\cite{Schuld2019,Park2020,Havlicek2019,Raghu2017,Schuld2021,Blank2020}. As with classical kernel methods, the goal of this embedding is to make the data easier to analyze. The kernel function is given by a pairwise similarity measure on the embedded data points (e.g., the fidelity of two embedded quantum states).  The kernel matrix is a matrix of all such pairwise similarities between input data points.
The quantum kernel can then be used with classical kernel machines, such as the support vector machine (SVM), in order to perform clustering, classification, or regression \cite{Abbas2020,Banchi2021,Canatar,Huang2021,Kubler2021,Park2020,Paine2022,Peters2022,Shaydulin,Shirai,Blank2020}. 

Because most learning properties can be inferred from the properties of the kernel matrix alone, quantum kernel methods are particularly amenable to theoretical analysis. In fact, many of the known speed-ups and improved generalization with fewer data points for quantum machine learning have been demonstrated using the quantum kernel approach~\cite{Huang2020,Liu2021,Huang2022}. The task-model alignment framework~\cite{canatar2021spectral} quantifies the number of samples a model needs to learn a target function. For each eigenmode of the kernel, the number of samples required to learn the mode depends on the corresponding eigenvalue. Therefore a ``flat'' spectrum implies a large sample complexity of learning each mode, making generalization near impossible for any target.

The task-model alignment perspective brings to the fore the central challenge of quantum machine learning, namely the need to introduce inductive bias into the quantum model to enable it to learn. K{\"u}bler et al.~\cite{Kubler2021} observe that without an inductive bias, the quantum model is guaranteed to be unable to generalize. Specifically, K{\"u}bler et al.~show that the spectrum of the kernel becomes ``flat'', and the largest eigenvalue becomes exponentially small with the number of qubits. Therefore the model requires exponentially many (with number of qubits) training points to learn even the first mode. As a consequence, for a quantum model to generalize efficiently, inductive bias must be introduced into the model \cite{Kubler2021,Cerezo2022,Cerezo2021a,Cerezo2021,Caro2022,Canatar,Bremner2015,Blank2020,Banchi2021,Arrasmith2022,Arrasmith,Abbas2020}.

Recently proposed approaches for introducing inductive bias into quantum machine learning models include projected kernels~\cite{Huang2021}, group-invariant machine learning~\cite{Larocca2022,2207.14413}, and quantum kernel bandwidth~\cite{Shaydulin,Canatar}. All three approaches prevent exponential ``flattening'' of the kernel spectrum and enable provable generalization in some settings. In projected or biased kernels, the quantum state generated by a feature map is projected onto a lower dimensional space of the full Hilbert space of the quantum state \cite{Huang2021}. 
However, appropriately selecting the feature map and projection for the kernel requires additional information about the target function that is, in most cases, unavailable. Previous results suggest that projected kernels are unlikely to provide advantage on classical data~\cite{Huang2021}. The group-invariant quantum machine learning approach has been proposed for quantum neural networks~\cite{1802.06002}, but applies trivially to quantum kernels as well~\cite{schuld2021supervised}. Specifically, problem-dependent symmetries are introduced into the feature map~\cite{Larocca2022}. However, this is only possible if such symmetries are present in the data. Finally, bandwidth tuning enables generalization by tuning a hyperparameter controlling the expressiveness of the model. This general approach has been shown to enable generalization on both synthetic and real data~\cite{Canatar} and does not depend on knowing any special property of the problem. In this paper, we focus on bandwidth tuning for improving the performance of quantum kernels.

Our main contribution is to provide strong numerical evidence that when quantum fidelity kernels generalize well, there are classical kernels that perform similarly well. Specifically, we tune the bandwidth of the quantum kernel so that the first eigenvalue of the kernel does not decay with the number of qubits, thereby ensuring that there exists at least one mode that can be learned with a number of training points that does not grow with the number of qubits. We then observe that the geometric difference between the quantum kernel and an appropriately constructed classical kernel decays rapidly as the number of qubits grows. Since a large geometric difference between quantum and classical kernels is required for quantum advantage~\cite{Huang2021}, our result suggests that quantum fidelity kernels are unlikely to provide an advantage on classical data without novel techniques for controlling the inductive bias.

The rest of the paper is organized as follows. In Section~\ref{sec:background}, we introduce the theoretical background underpinning our numerical experiments. We introduce our numerical experiment design as well as the feature maps and data distributions that we study in Section~\ref{sec:methods}. In Section~\ref{sec:experiments}, we present the results of our numerical experiments. In Section~\ref{sec:discussion}, we discuss the context and the limitations of the numerical experiments presented in this work and suggest future research directions.

\section{\label{sec:background} Background}

We begin by briefly reviewing the relevant concepts. We study the problem of classification. Given a training set $\{\bx_i,y_i\}_{i=1}^P$ with $\bx$ drawn from some marginal probability distribution on $\mathcal{X}\subset \mathbb{R}^n$ and $\bar{f}(x_i) = y_i$, the goal is to learn the target function $\bar{f}(x)$. In our numerical experiments with real data, we focus specifically on the problem of binary classification, where $y_i\in\{-1,+1\}$, though our results apply to quantum supervised learning in general.

For an $n$-qubit quantum system, we consider quantum feature maps $\psi$ that map a data point $\bx$ drawn from some distribution defined by a probability density function $p:\mathcal{X}\rightarrow\mathbb{R}$ to an $n$-qubit quantum state given by a density matrix $\ket{\psi(\bx)}\bra{\psi(\bx)}=\rho(\bx)$. We denote by $\mathcal{H}$ the quantum Hilbert space of all density matrices encoding $n$-qubit quantum states endowed with inner product $\langle \rho, \rho'\rangle_{\mathcal{H}} = \Tr(\rho\rho')$. We can interpret $\mathcal{H}$ as a finite-dimensional reproducing kernel Hilbert space (RKHS) associated with the kernel $k(\bx, \bx') = \langle \psi(\bx), \psi(\bx')\rangle_{\mathcal{H}} = \Tr(\rho(\bx)\rho(\bx'))$. Then for a set of data points $\{\bx_j\}_{j=1}^P$ the kernel matrix is defined element-wise as $K_{ij}=k(\bx_i, \bx_j)$. Its continuous analogue is the integral kernel operator $T_k: L_2(\mathcal{X}) \rightarrow L_2(\mathcal{X})$ defined according to its action:
\begin{equation}\label{eq:integral_op}
    (T_k f)(\bx) = \int_{\mathcal{X}} k(\bx, \bx') f(\bx') p(\bx) d\bx,
\end{equation}
with orthonormal eigenfunctions $T_k \phi_k = \eta_k \phi_k$ from Mercer's theorem~\cite{scholkopf2002learning}. While many theoretical results are formulated in terms of the eigenvalues of the integral operator (e.g., \cite{Kubler2021}), the eigenvalues $\gamma_j$ of the kernel matrix $K$ approach the corresponding eigenvalues $\eta_j$ of the integral operator in the limit of infinite number of data points. Therefore in numerical experiments we focus on studying the spectrum of the kernel matrix $K$.

For quantum advantage in machine learning, two necessary (but not sufficient) conditions must hold. First, the quantum model must generalize well. The framework of task-model alignment~\cite{canatar2021spectral}, which is based on the replica method of statistical physics~\cite{sompolinsky1992examples, sompolinsky1999statistical, mezard2009information,advani2013statistical}, allows us to connect the spectral properties of the kernel to model generalization. Informally, the target function can be decomposed in the eigenbasis of the integral operator $T_k$. Then the model needs to ``learn'' a sufficient number of eigenmodes of $T_k$ so that the cumulative weight of the target on these modes is close to one. The cost of learning each mode is dependent on the corresponding eigenvalue $\eta$. As a consequence, if the largest eigenvalue $\eta_{\max}$ is small, even the first mode cannot be learned, and generalization is impossible.  A simplified version of this observation is given in Ref.~\cite[Theorem 3]{Kubler2021}, wherein for a fixed target function the number of data points needs to scale as $\frac{1}{\sqrt{\gamma_{\max}}}$ to keep the kernel ridge regression error constant.  Interested readers are referred to Refs.~\cite{canatar2021spectral,Canatar,Kubler2021} for a more detailed discussion.

Second condition is that the quantum model must be ``sufficiently different'' from \emph{all} classical models. Consider a quantum model that is linear in the (quantum) feature space (i.e., $f(x)=Tr(O\rho(x))$). This class of models includes quantum kernel methods.
In Huang et al.~\cite{Huang2021}, the authors demonstrate that the prediction error for the quantum model and trained quantum/classical kernel model $h(x)$ can be upper bounded by the trained model's complexity $s_K(N)$. The simplified prediction error bound for a new test sample x is given by:
\begin{equation}
\mathbb{E}_x=|h(x)-f(x)|\leq c \sqrt{\frac{s_K(N)}{N}} \sim \mathcal{O}\left(\ \sqrt{\frac{s_K(N)}{N}} \right)\
\label{eq:pred_bound}
\end{equation}
for $c > 0$ and $s_K(N)$ defined as:

\begin{equation}
s_K(N)=\sum_i^N \sum_j^N (K^{-1})_{ij}Tr(O\rho(x_i))Tr(O\rho(x_j))
\end{equation}

Furthermore, the authors introduce the asymmetric quantity, the geometric difference $g_d$:

\begin{equation}
    g_d(K_1\|K_2)=\sqrt{\| \sqrt{K_2} K_1^{-1} \sqrt{K_2} \|_{\infty}}
\end{equation}
which quantifies the difference between two kernel matrices and is independent of data labels. The trained model complexities can then be shown to satisfy the inequality $s_{K_1} \leq g(K_1 \| K_2)^2 s_{K_2}$ when $s_{K_1} \leq \ s_{K_2}$. Thus, for a quantum kernel, $K_Q$ to significantly outperform a classical kernel $K_C$ on a dataset it is necessary but not sufficient that $g_d(K_C \| K_Q) \ll \sqrt{N}$. 
To satisfy the necessary condition, the geometric difference needs to be minimized over many optimized classical machine learning models. However, to show that this condition does not hold, it is sufficient to present one classical model with a small geometric difference, which is the approach we take in this work. 

To solve the classification problem and obtain generalization error on real data, we use the quantum kernel matrix in a support vector machine (SVM). SVM aims to find a separating hyperplane maximizing the distance to any data point in either class (margin)~\cite{vapnik1995nature}. The hyperplane is obtained by solving the following convex optimization problem 

\begin{equation}
     \sum_{i=1}^{N} \alpha_{i}-\frac{1}{2} \sum_{i=1}^{N} \sum_{j=1}^{N} \alpha_{i} \alpha_{j} y_{i} y_{j} k(\mathbf{x}_{i}, \mathbf{x}_{j}),
     \label{eq:SVC_dual_obj}
 \end{equation}
 subject to $\sum_{i=1}^{N} \alpha_{i} y_{i}=0$ and $0\leq\alpha_i\leq C$, $i=1,\ldots,N$, where $C$ is a regularization hyperparameter.

\section{\label{sec:methods} Methods}

To numerically investigate the possibility of quantum advantage with quantum kernels, we study the scaling behavior with the number of qubits for kernels that are guaranteed to generalize on at least some target with a fixed number of training points. To enforce this condition, we follow 
Ref.~\cite{Shaydulin} and perform hyperparameter tuning on the quantum feature maps. For the real datasets, we tune the hyperparameters in order to maximize the model test score. For the unlabeled synthetic data sets, the test score is not available. We therefore tune the hyperparameters in order to keep the maximum kernel matrix eigenvalue fixed at a target eigenvalue with increasing system size in our simulations, which guarantees that the model is able to generalize with a constant (with number of qubits) number of training data points on at least one target function. Next, we obtain the geometric difference $g_d$ between the quantum fidelity kernel matrix and a classical kernel matrix computed on the same data distribution. Specifically, we perform a grid search on the hyperparameter of our classical kernel in order to minimize geometric difference between the quantum and classical kernel matrices. We now discuss our choice of datasets and quantum feature maps. %

\subsection{\label{sec:maps}Feature Maps}

We consider two distinct families of feature maps in our numerical experiments. We consider a feature map in Eq.~\ref{iqp} inspired by the instantaneous quantum polynomial (IQP) feature map studied in Ref.~\cite{Havlicek2019}.

\begin{equation}
    \ket{\psi(x_i)} = \hat{U}_Z(x_i)\hat{H}^{\otimes d}\hat{U}_Z(x_i)\hat{H}^{\otimes d}\ket{0}
    \label{iqp}
\end{equation}

where

\begin{equation}
   \hat{U}_Z(x_i)=\exp{( \sum_{j=1}^{d}{\lambda x_{ij} \hat{\sigma}_j^z} + \sum_{j,k=1}^{d}{\lambda^{\alpha} x_{ij} x_{ik} \hat{\sigma}_j^z \hat{\sigma}_k^z} )}.
    \label{Uz}
\end{equation}

Our IQP style feature map has a single hyperparameter, the scaling factor $\lambda$. In our numerical experiments, the scaling factor $\lambda$ will be tuned to control the eigenvalue spectrum of our kernel matrix. The feature map also has the parameter $\alpha$ which we keep fixed during hyperparameter tuning. We study feature maps with $\alpha\in \{0.5,1.0,2.0\}$. When $\alpha=2.0$, the feature map is identical to that studied in Eq. 4 of Ref.~\cite{Shaydulin}. In this case, the scaling factor $\lambda$ can be seen as re-scaling the data points $x_i \rightarrow \lambda x_i$ and the scaling factor behaves like bandwidth in classical kernel methods \cite{Shaydulin}.

We also consider a feature map using 1D Heisenberg interactions

\begin{equation}
    \ket{\psi(x_i)}= \Bigl( \prod_{j=1}^{d}\exp{\bigl( -i \hat{H}(x_{ij}) \bigr)} \Bigr) ^{n_l} \ket{0},
    \label{hfeaturemap}
\end{equation}
where
\begin{equation}
    \hat{H}(x_{ij})=\Vec{S}_j  \cdot \Vec{S}_{j+1} +\lambda x_{ij} \sigma^z_j.
    \label{heq}
\end{equation}

In the Heisenberg feature map, we again have a single hyperparameter, the scaling factor $\lambda$. Throughout our numerical experiments, we fix the number of layers in Eq.~\ref{hfeaturemap}, $n_l$, to 4.

In our numerical experiments, we choose our classical kernel to be a 'classical' version of our quantum fidelity kernel. In the case of the IQP style feature map, we compare our quantum fidelity kernel with a fidelity kernel produced by the feature map
\begin{equation}
    \ket{\psi(x_i)} = \hat{U}_Z(x_i)\hat{H}^{\otimes d}\ket{0}.
    \label{classical_iqp}
\end{equation}

Note that while the output distribution of this feature map is believed to be hard to reproduce classically the fidelity kernel produced by the inner product of the feature map is classically simulatable in polynomial time \cite{Bremner2015,Havlicek2019}. In the case of Heisenberg feature map, we compare to a product state feature map wherein we have removed the Heisenberg interaction term:
\begin{equation}
    \ket{\psi(x_i)}= \prod_{j=1}^{d}\exp{\bigl( -i \lambda x_{ij} \sigma^z_j \bigr)} \ket{0}
    \label{classical_hfeaturemap}
\end{equation}

We choose these ``classical versions'' of the kernels \eqref{iqp}, \eqref{hfeaturemap} because they are reasonable ansatze for minimizing the geometric difference between the quantum fidelity kernel and a classical kernel. However, other classical kernels, including more sophisticated neural network architectures could also have been chosen that may have recovered the same or smaller geometric differences. For all feature maps studied the number of qubits is equal to the data point dimension.

\subsection{\label{sec:data} Data Distributions}

As outlined in Sec.~\ref{sec:background}, a large task-model alignment indicates that a kernel method will be able to generalize efficiently to new data \cite{canatar2021spectral}. A prerequisite for this is that the eigenvalue spectrum of the kernel must not be ``flat'', that is the largest eigenvalue must not be ``too small''. Concretely, for the number of training data points required to obtain constant generalization error with the number of qubits, the largest eigenvalue must not decrease. In our analysis, we examine the largest eigenvalue of our kernel matrix and the geometric difference between a quantum and classical kernel. Our analysis is agnostic to any accompanying labels to data points. We examine both real data distributions and synthetically generated unlabeled data distributions. For real datasets, we use the Plasticc dataset and the Fashion-Mnist dataset.
Our synthetic data distributions are generated by randomly sampling data elements $x_{ij}$ in our data points $x_i$ from the symmetric generalized normalized distribution
\begin{equation}
    x_{ij} \sim \frac{\beta}{2\sigma\Gamma(1/\beta)}\exp{\left(-\left|\frac{x-\mu}{\sigma}\right|^{\beta}\right)},
    \label{gennorm}
\end{equation}
where $\mu=0$ and $\sigma=1$. After sampling all data points for a data distribution, the data elements are standardized and normalized. In our experiments, we study distributions with $\beta \in \{0.1, 1.0, 2.0\}$. In the case of $\beta=1.0$ and $2.0$ the distribution reduces to the laplacian and the normal distribution respectively. The effect of $\beta$ on the distribution is that smaller values of $\beta$ lead to larger kurtosis of the distribution (kurtosis = $\frac{\Gamma(5/\beta)\Gamma(1/\beta)}{\Gamma(3/\beta)^2} -3$).

We also consider a correlated data distribution using the symmetric generalized normalized distribution. In this case, the correlated data elements are generated using
\begin{equation}
    \begin{split}
    & x_{ij} \sim r \cdot x_{i(j-1)} + \\ 
    & \sqrt{1-r^2} \cdot \frac{\beta}{2\sigma\Gamma(1/\beta)}\exp{\left(-\left|\frac{x-\mu}{\sigma}\right|^{\beta}\right)},
    \end{split}
    \label{correlated_gennorm}
\end{equation}
where the expected correlation between the data element $x_{ij}$ and $x_{i(j-1)}$ is r.

\section{\label{sec:experiments} Numerical Experiments}

We now present the numerical evidence that quantum advantage with quantum fidelity kernel methods is unlikely on classical data. We observe that as the number of qubits grows, the quantum model behaves ``more classically'' as evidenced by the decreasing geometric difference with classical kernels. For the Plasticc and Fashion-Mnist data sets, the decrease holds both for fixed classical kernel and the best-performing one. We denote the geometric difference between the classical and quantum kernels, $g_d(K_C \| K_Q)$, as simply $g_d$.

\subsection{\label{sec:iqp_experiments} IQP Style Feature Map}
We first present the numerical results for the IQP style feature map (Eq.~\ref{iqp}) and the classical fidelity kernel generated by the feature map in (Eq.~\ref{classical_iqp}) for $\alpha \in \{0.5, 1.0, 2.0\}$. We present results for the Plasticc and Fashion-Mnist data sets and for synthetic data sets. All kernel matrices use N=500 randomly sampled data points from the uncorrelated data distribution (Eq.~\ref{gennorm}) unless explicitly stated otherwise.

\subsubsection{\label{sec:plasticc} $\alpha = 2.0,$ \hspace{2pt} Real Datasets}

\begin{figure*}[]

\begin{tabular}{c}

\subfloat[]{
\includegraphics[width=132pt]{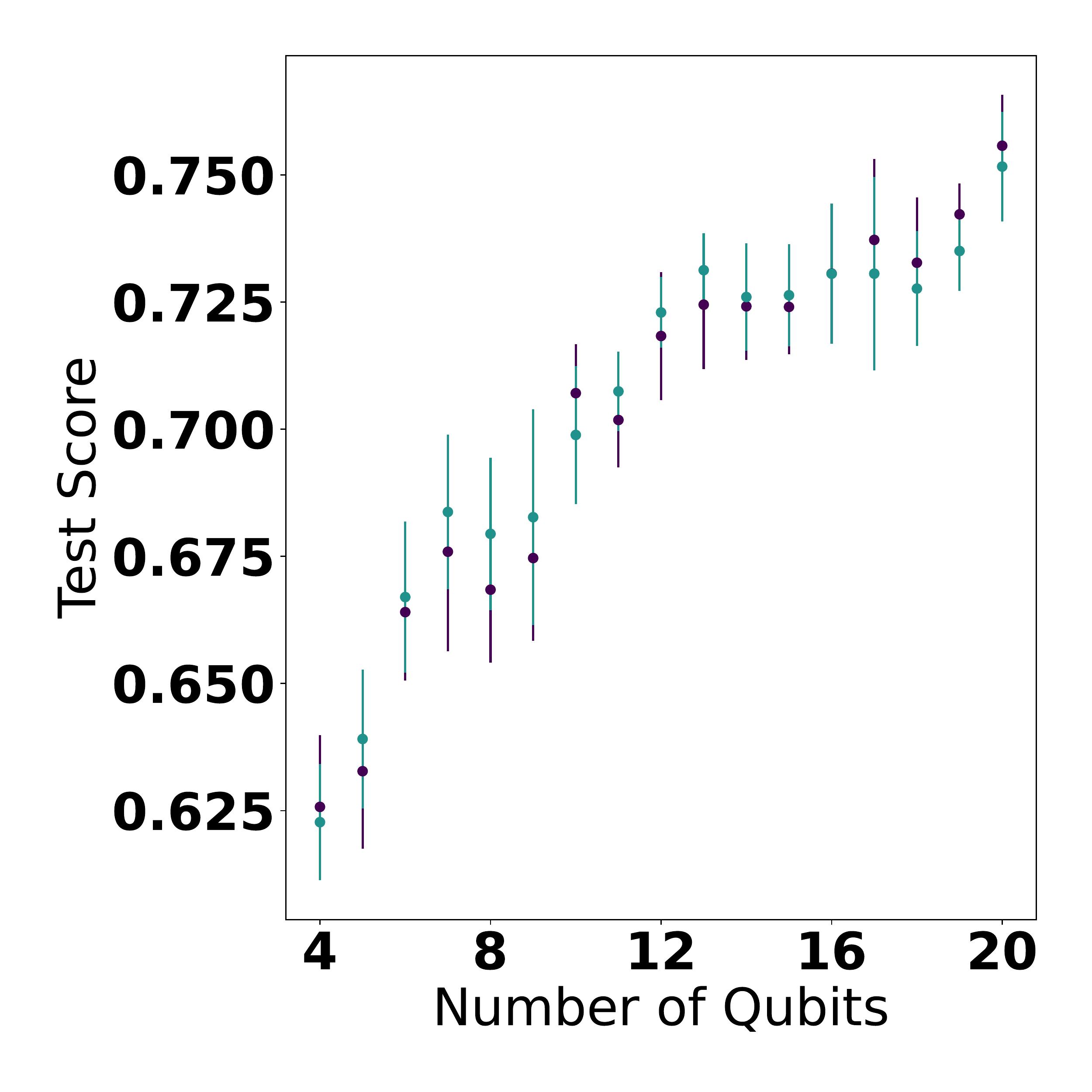}
\label{fig:real_data-a}
}
\hspace{-20pt}
\subfloat[]{
\includegraphics[width=132pt]{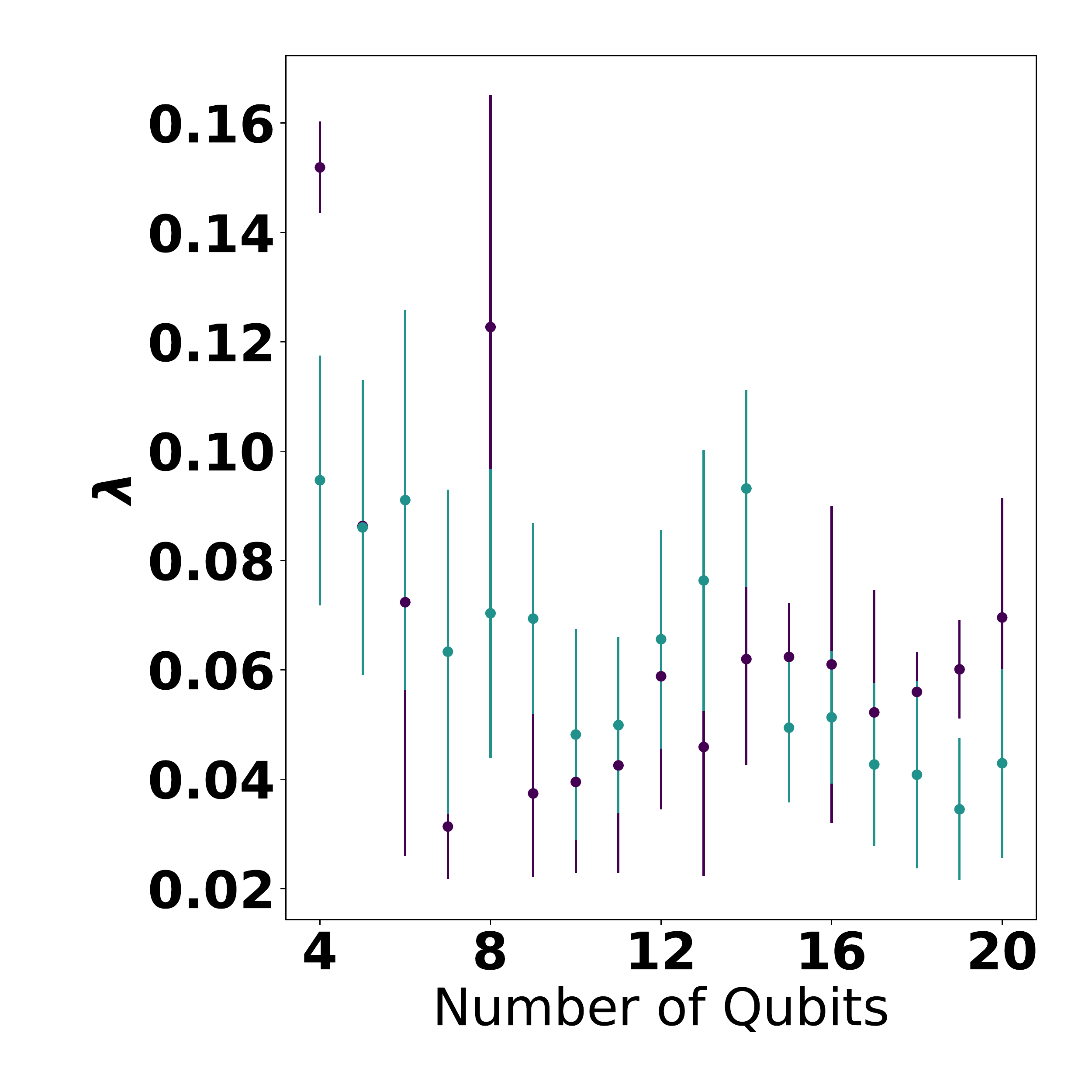}
\label{fig:real_data-b}
}
\hspace{-20pt}
\subfloat[][]{
\includegraphics[width=132pt]{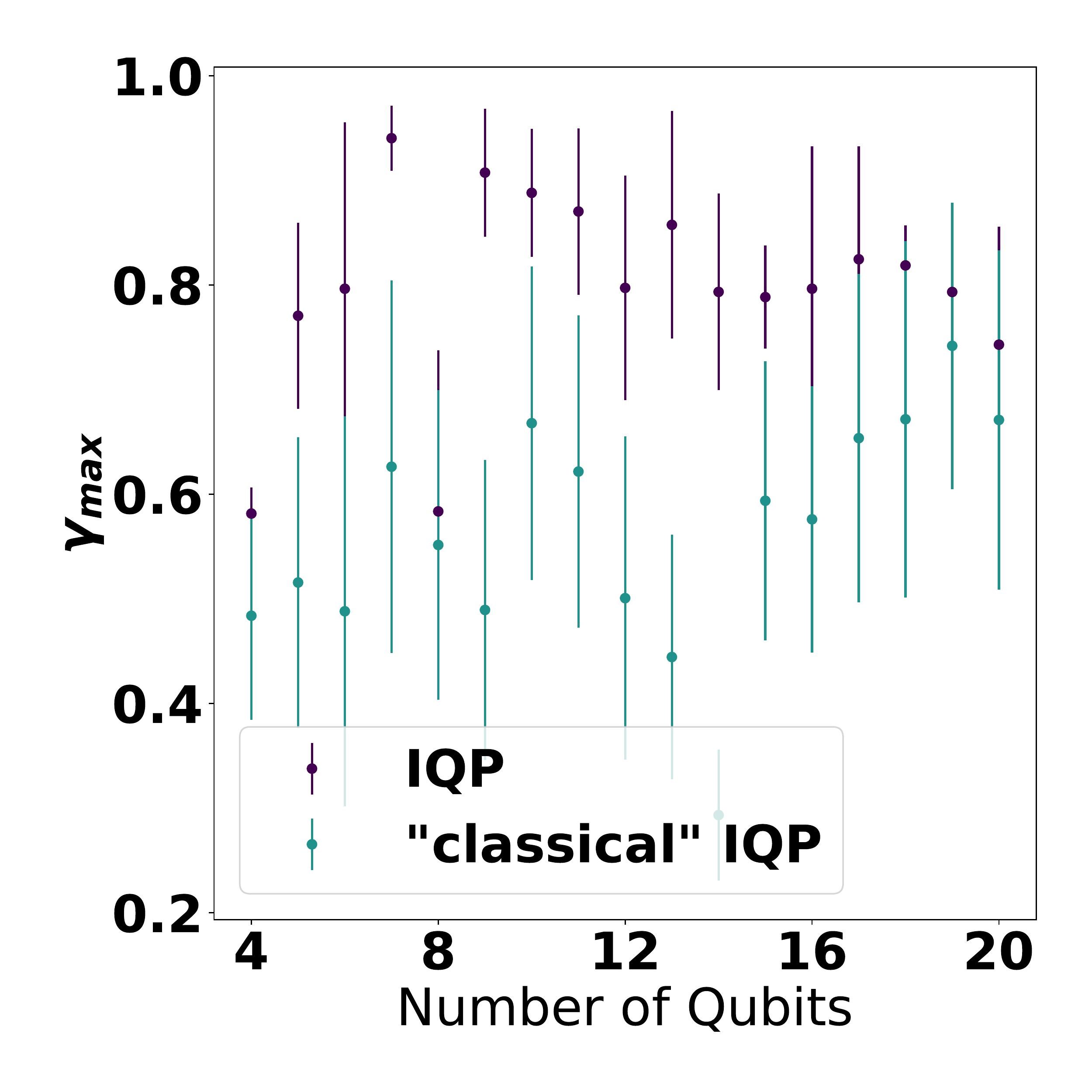}
\label{fig:real_data-c}
}
\hspace{-20pt}
\subfloat[]{
\includegraphics[width=135pt]{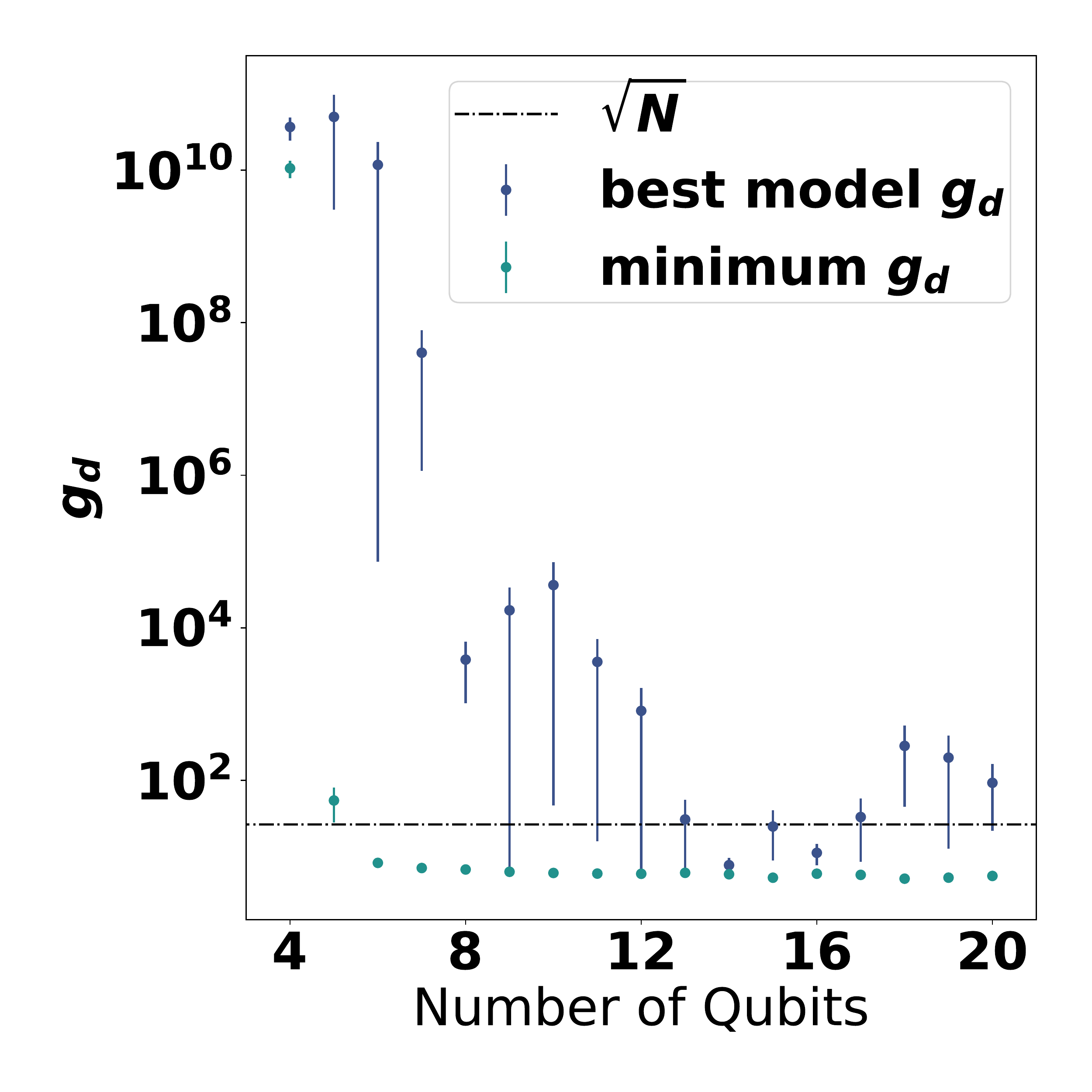}
\label{fig:real_data-d}
}
\end{tabular}
\caption{For the Plasticc dataset, we plot the test score \protect\subref{fig:real_data-a}, scaling factor $\lambda$ \protect\subref{fig:real_data-b}, maximum eigenvalue $\gamma_{max}$ \protect\subref{fig:real_data-c} and geometric difference \protect\subref{fig:real_data-d} vs. the number of qubits for both the quantum and 'classical' IQP style feature map with $\alpha=2.0$. For all for plots, we show results for the highest test score kernels. In \protect\subref{fig:real_data-d}, we also include the minimum $g_d$ found between the highest test score quantum kernel and a grid search over classical kernels with varying $\lambda$.}
\label{fig:real_data}
\end{figure*}

\begin{figure*}[]
\centering
\begin{tabular}{c}

\subfloat[]{
\includegraphics[width=172pt]{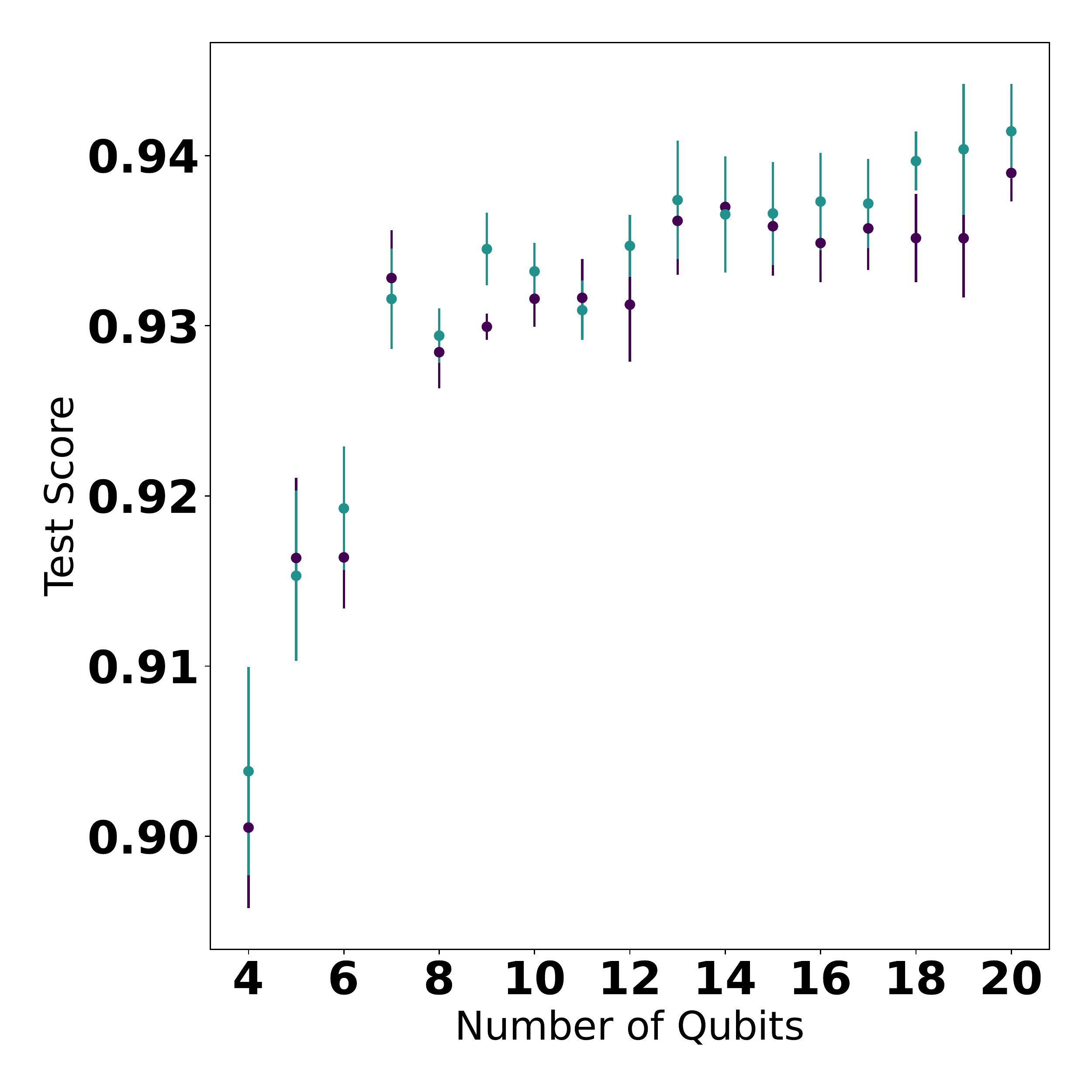}
\label{fig:fm_data-a}
}
\hspace{-20pt}
\subfloat[]{
\includegraphics[width=172pt]{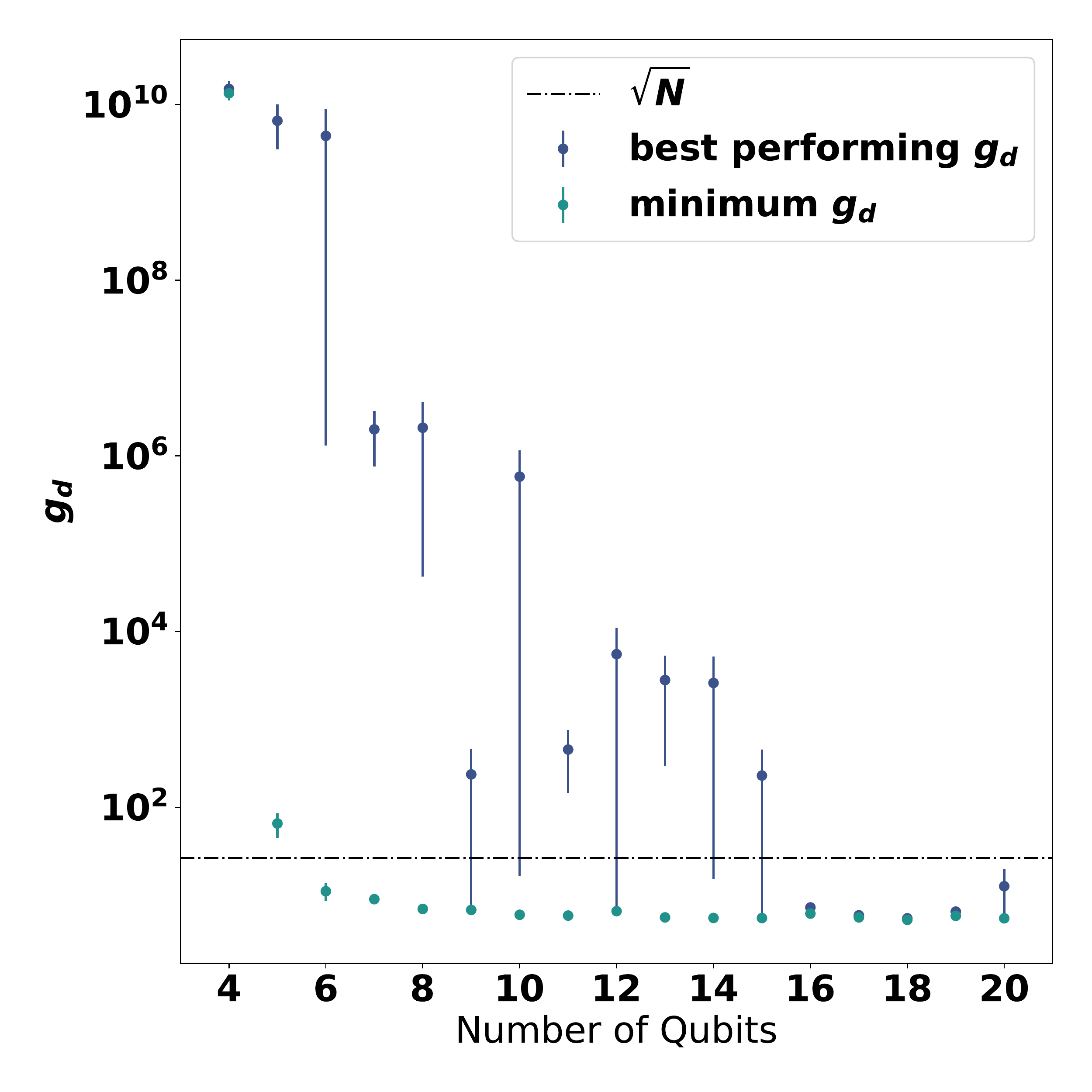}
\label{fig:fm_data-b}
}
\hspace{-20pt}
\subfloat[]{
\includegraphics[width=172pt]{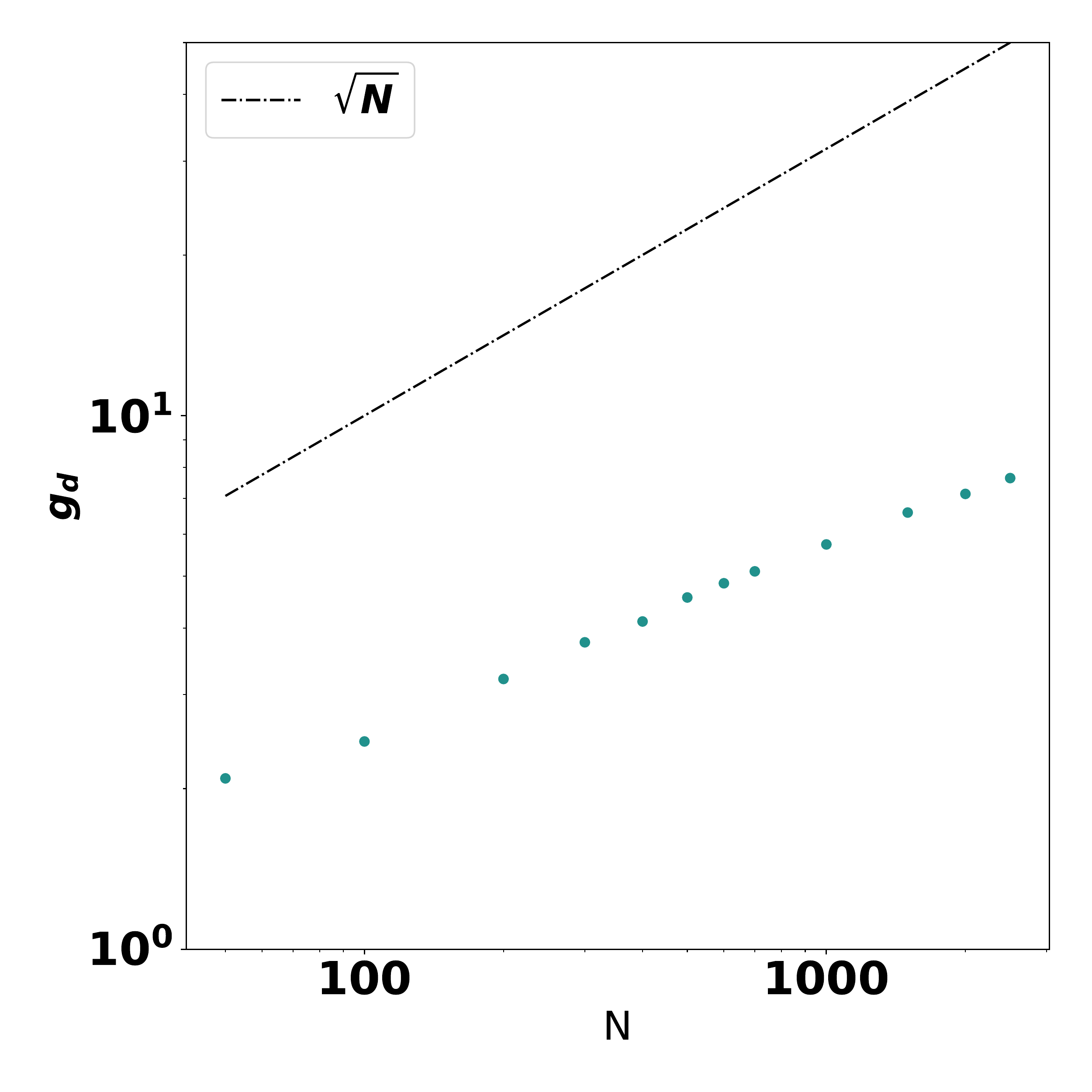}
\label{fig:fm_data-c}
}
\end{tabular}
\caption{For the Fashion-Mnist dataset, we plot the test score \protect\subref{fig:fm_data-a}, geometric difference vs. the number of qubits \protect\subref{fig:fm_data-b} and vs. the number of data points \protect\subref{fig:fm_data-c}  for both the quantum and 'classical' IQP style feature map with $\alpha=2.0$. In \protect\subref{fig:fm_data-a}, plot the test scores of the two top performing models. In \protect\subref{fig:fm_data-b}, plot the $g_d$ between the two high test score kernels as well as the minimum $g_d$ found between the highest test score quantum kernel and a grid search over classical kernels with varying $\lambda$. In \protect\subref{fig:fm_data-c}, we plot $g_d$ vs. the number of data points for 16 qubit feature maps. The feature maps use the $\lambda$s that minimize $g_d$ for N=700.}
\label{fig:fm_data}
\end{figure*}

In this section, we present results for the IQP style feature map with $\alpha=2.0$ applied to the Plasticc data set and the Fashion-Mnist data. We sample from the Plasticc dataset 700 training vectors and 200 test vectors. For the Fashion-Mnist data set, we sample 700 training vectors and 200 test vectors in addition to varying the number of training vectors in Fig.~\ref{fig:fm_data-c}. For number of qubits 4 to 20, we perform principal component analysis (PCA) on the data points and reduce the data set dimension to the number of qubits in our feature map before performing standardization and normalization. For both the IQP style feature map (Eq. \ref{iqp}) and the 'classical' IQP feature map (Eq.~\ref{classical_iqp}), we tune the scaling factor, $\lambda$, and the SVM regularization parameter, C in order to find the best performing SVM on the test data for each feature map. In Fig.~\ref{fig:real_data}, we show the results for the best performing feature maps for each system size studied.

For the Plasticc data set, we plot four key quantities as a function of the number of qubits: test score, scaling factor $\lambda$, maximum kernel eigenvalue $\gamma_{max}$ and the geometric difference, $g_d$. We observe that including more principal components increases the test score performance of both the quantum and classical fidelity kernel  (Fig.~\ref{fig:real_data}\subref{fig:real_data-a}). The IQP style feature map may perform marginally better than the specific classical model we compare it to and have larger $\gamma_{max}$ (Fig.~\ref{fig:real_data}\subref{fig:real_data-c}). In Fig.~\ref{fig:real_data-d}, we plot $g_d$ against the number of qubits for between the highest test score kernels for both the classical and quantum IQP style feature map. In addition, we plot the minimum $g_d$ between the highest test score quantum kernel and a grid search over the classical feature map scaling factor. We observe a decay in the geometric difference $g_d$ with increasing number of qubits (Fig.~\ref{fig:real_data}\subref{fig:real_data-d}). While we make no definitive claims about the future trend of $g_d$, for the system sizes we study, the $g_d$ between the two top-performing kernels approaches $\sqrt{N}$. This indicates, for the data distribution (independent of the labels), the two models can still differ in their predictions. However, the minimum $g_d$ found between the models indicates that the potential for quantum advantage decreases with system size.

For the Fashion-Mnist data set, we plot two key quantities vs. the number of qubits. As in the Plasticc data set, we plot the test score and $g_d$ vs. the number of qubits. In addition, we plot $g_d$ vs. the number of training vectors. Again, we observe that including more principal components increases the test score of both the quantum and classical fidelity kernel (Fig.~\ref{fig:fm_data-a}). For this data set, the ``classical'' IQP fidelity kernel performs better than the quantum kernel on the test set. As before, we observe the decay of $g_d$ with increasing number of qubits (Fig.~\ref{fig:fm_data-b}). For this data set, the $g_d$ between the top kernels and the minimum $g_d$ found via grid search over the classical feature map scaling factor both decay below $\sqrt{N}$. The decay in $g_d$ indicates the potential for the compared models to make different predictions decreases with system size. In addition to decay with the number of qubits, we observe that the minimum $g_d$ grows slower than $\sqrt{N}$, suggesting that quantum advantage is unlikely. Fig.~\ref{fig:fm_data-c} plots $g_d$ against the number of training vectors for a 16 qubit system.

\subsubsection{\label{sec:fixed_alpha_beta} $\alpha = 2.0, \hspace{2pt} \beta=1.0$ uncorrelated data}

\begin{figure*}[]
\centering
\begin{tabular}{c}

\subfloat[]{
\includegraphics[height=120pt]{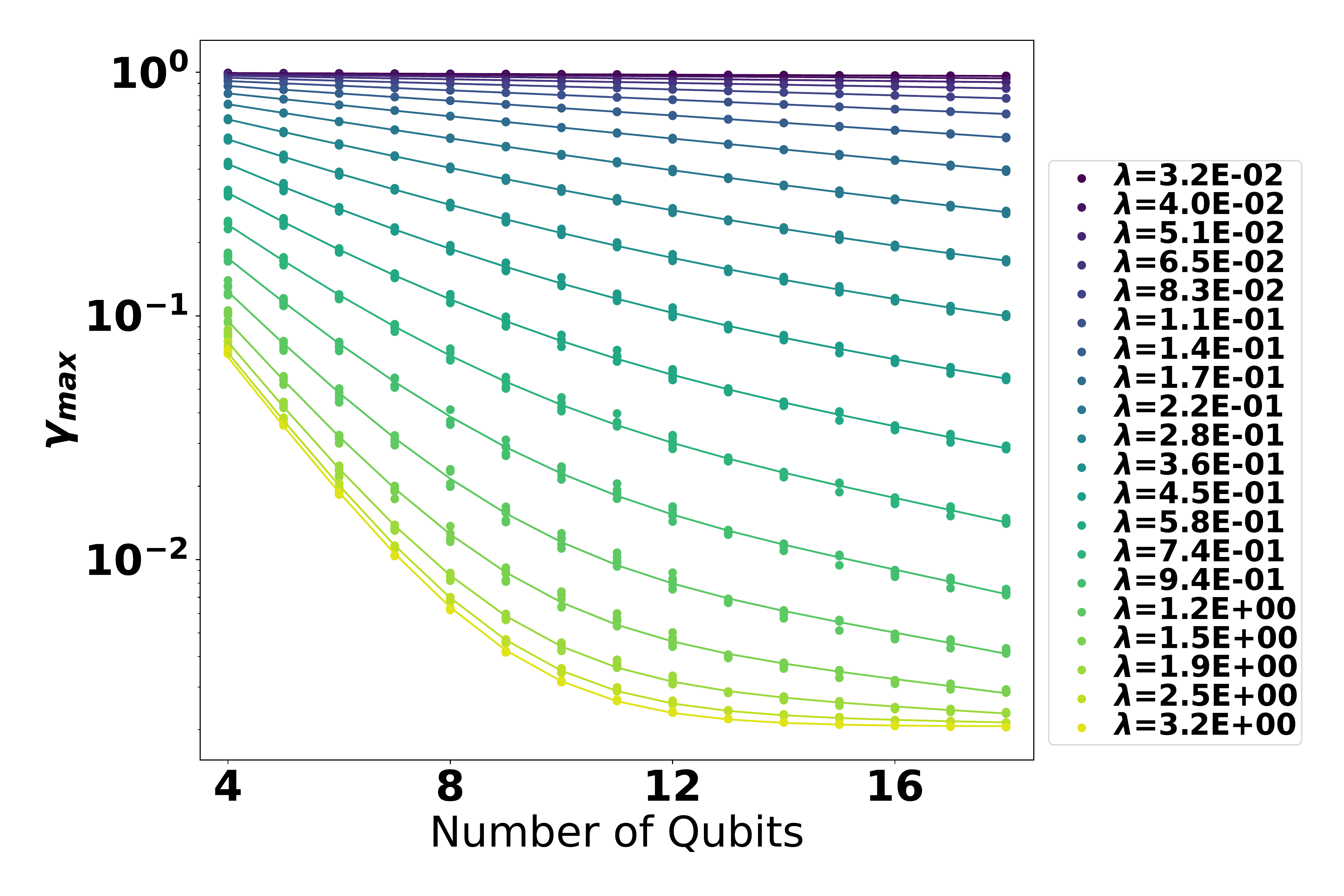}
\label{fig:fixed_ab_curves-a}
}
\hspace{-20pt}
\subfloat[]{
\includegraphics[height=120pt]{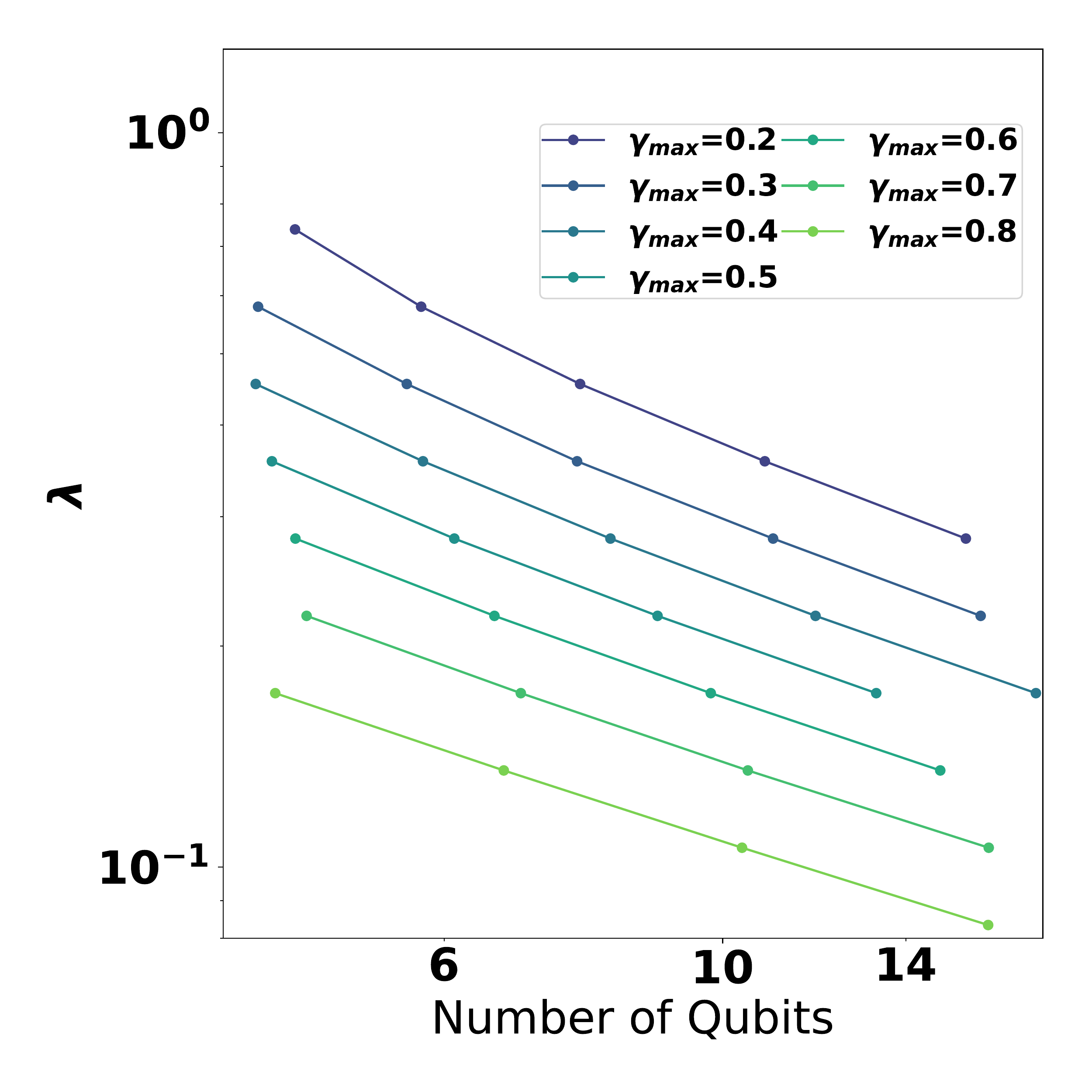}
\label{fig:fixed_ab_curves-b}
}
\hspace{-20pt}

\subfloat[]{
\includegraphics[height=120pt]{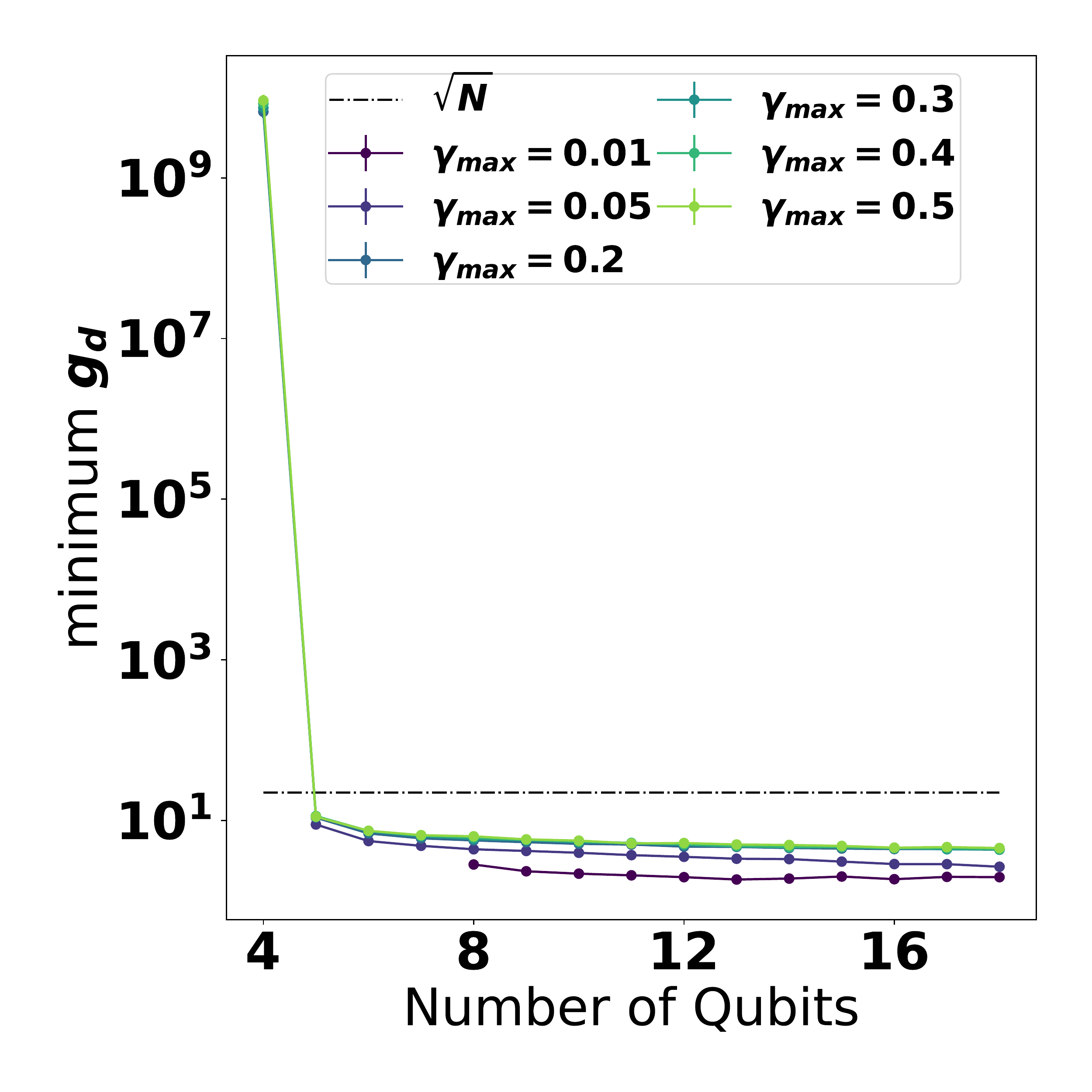}
\label{fig:fixed_ab_curves-c}
}
\hspace{-20pt}

\subfloat[]{
\includegraphics[height=120pt]{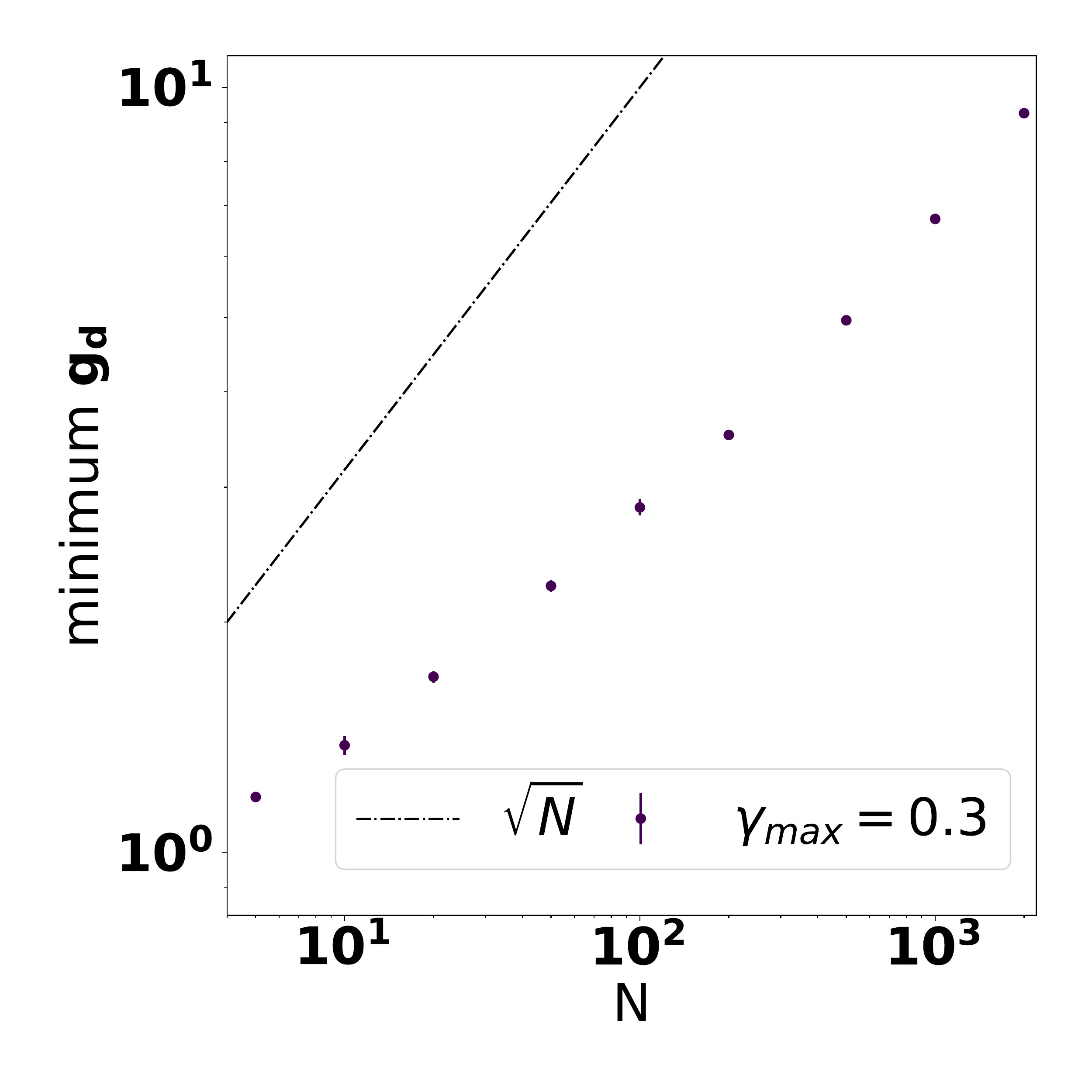}
\label{fig:fixed_ab_curves-d}
}
\end{tabular}
\caption{Plots for $\alpha=2.0$ and $\beta=1.0$ using the IQP style feature map. \protect\subref{fig:fixed_ab_curves-a} Maximum kernel matrix eigenvalue vs. number of qubits for varying $\lambda$. \protect\subref{fig:fixed_ab_curves-b} $\lambda$ vs number of qubits for a target $\gamma_{max}$. \protect\subref{fig:fixed_ab_curves-c}  Minimum geometric difference, $g_d$, vs. number of qubits for varying $\gamma_{max}$. \protect\subref{fig:fixed_ab_curves-d} Minimum geometric difference, $g_d$, vs. number of data points, N, in distribution for $\gamma_{max}=0.3$ and a 12 qubit feature map. A geometric difference of $O(\sqrt{N})$ (the black dashed line) is necessary for any potential quantum advantage to exist.}
\label{fig:fixed_ab_curves}
\end{figure*}

First, for $\alpha=2.0$ and $\beta=1.0$, we perform a grid search with our IQP feature map for $10^{-1.5} \leq \lambda \leq 10^{0.5}$ and number of qubits 4 to 18. In Fig.~\ref{fig:fixed_ab_curves}\subref{fig:fixed_ab_curves-a}, we plot the maximum kernel eigenvalue as a function of the number of qubits for varying $\lambda$. The solid lines represent interpolations between the mean values of the maximum kernel eigenvalue. The maximum kernel eigenvalue rapidly decays for fixed $\lambda$. For the largest $\lambda$s we observe the maximum kernel eigenvalue decay to $1/N=1/500$ indicating a completely flat eigenvalue spectrum. In Fig.~\ref{fig:fixed_ab_curves}\subref{fig:fixed_ab_curves-b}, we plot $\lambda$ vs. number of qubits for fixed maximum kernel eigenvalues. The points for this plot are extracted from the grid-search interpolations in Fig.~\ref{fig:fixed_ab_curves}\subref{fig:fixed_ab_curves-a}.

Using the interpolation in Fig.~\ref{fig:fixed_ab_curves}\subref{fig:fixed_ab_curves-b}, we hyperparameter tune $\lambda$ in order to keep $\gamma_{max}$ approximately constant with increasing number of qubits. In practice, we find that $\gamma_{max}$ remains within 0.08 and 20\% of our target for all kernel matrices studied in this paper. We compare the quantum fidelity kernel matrices to classical fidelity kernel matrices using the feature map in Eq.~\ref{classical_iqp} with the same number of qubits and the same randomly sampled data points. For the classical fidelity kernels, we perform a grid search over $10^{-1.5} \leq \lambda \leq 10^{0.5}$ in an attempt to minimize the distance between the classical fidelity kernel matrix and our quantum fidelity kernel matrix. In Fig.~\ref{fig:fixed_ab_curves}\subref{fig:fixed_ab_curves-c}, we plot the results of the grid searches. We plot the minimum $g_d$ found vs. number of qubits for $\gamma_{max} \in \{0.01, 0.05, 0.2, 0.3, 0.4, 0.5\}$. For all $\gamma_{max}$, we observe a rapid decay in $g_d$ with increasing number of qubits to well below $\sqrt{N}$ indicating that there can be no quantum advantage for this data point distribution and feature map combination. This is further supported by the observation that the geometric difference grows slower than $\sqrt{N}$. Specifically, in Fig.~\ref{fig:fixed_ab_curves}\subref{fig:fixed_ab_curves-d}, we repeat our analysis using $\gamma_{max}=0.3$ and a 12 qubit system while varying the number of data points in our distribution (and thus the kernel matrix size) N. The geometric difference $g_d$ remains significantly below the $\sqrt{N}$ line and grows at a slower trend, indicating that no quantum advantage can exist.

\subsubsection{\label{sec:vary_alpha_beta} $\alpha \in \{0.5, 1.0, 2.0\}, \hspace{2pt} \beta \in \{0.1, 1.0, 2.0\}$ uncorrelated data}

\begin{figure*}
\centering
\begin{tabular}{c}

\subfloat[]{
\includegraphics[width=132pt]{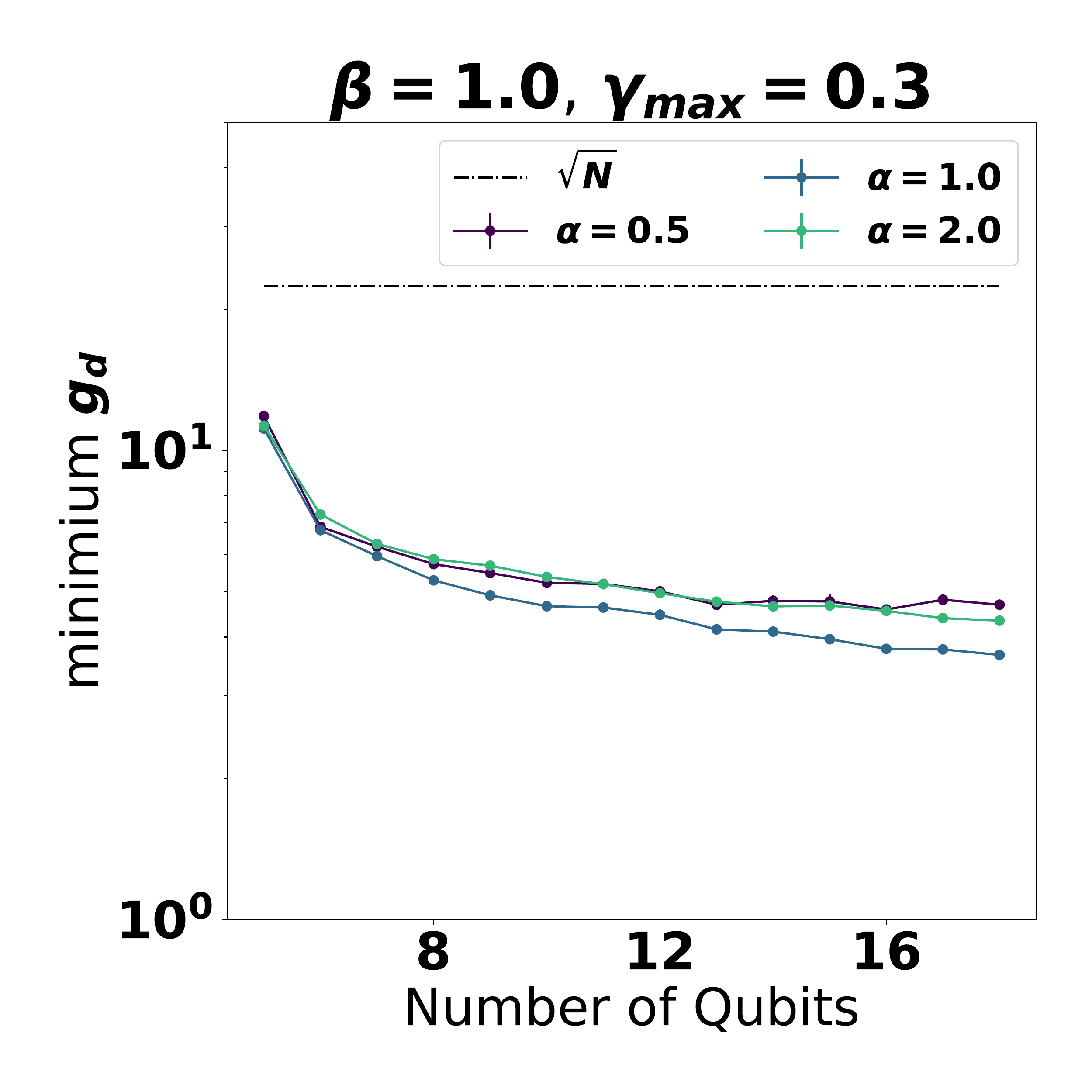}
\label{fig:vary_ab_gd_scaling-a}
}
\hspace{-20pt}
\subfloat[]{
\includegraphics[width=132pt]{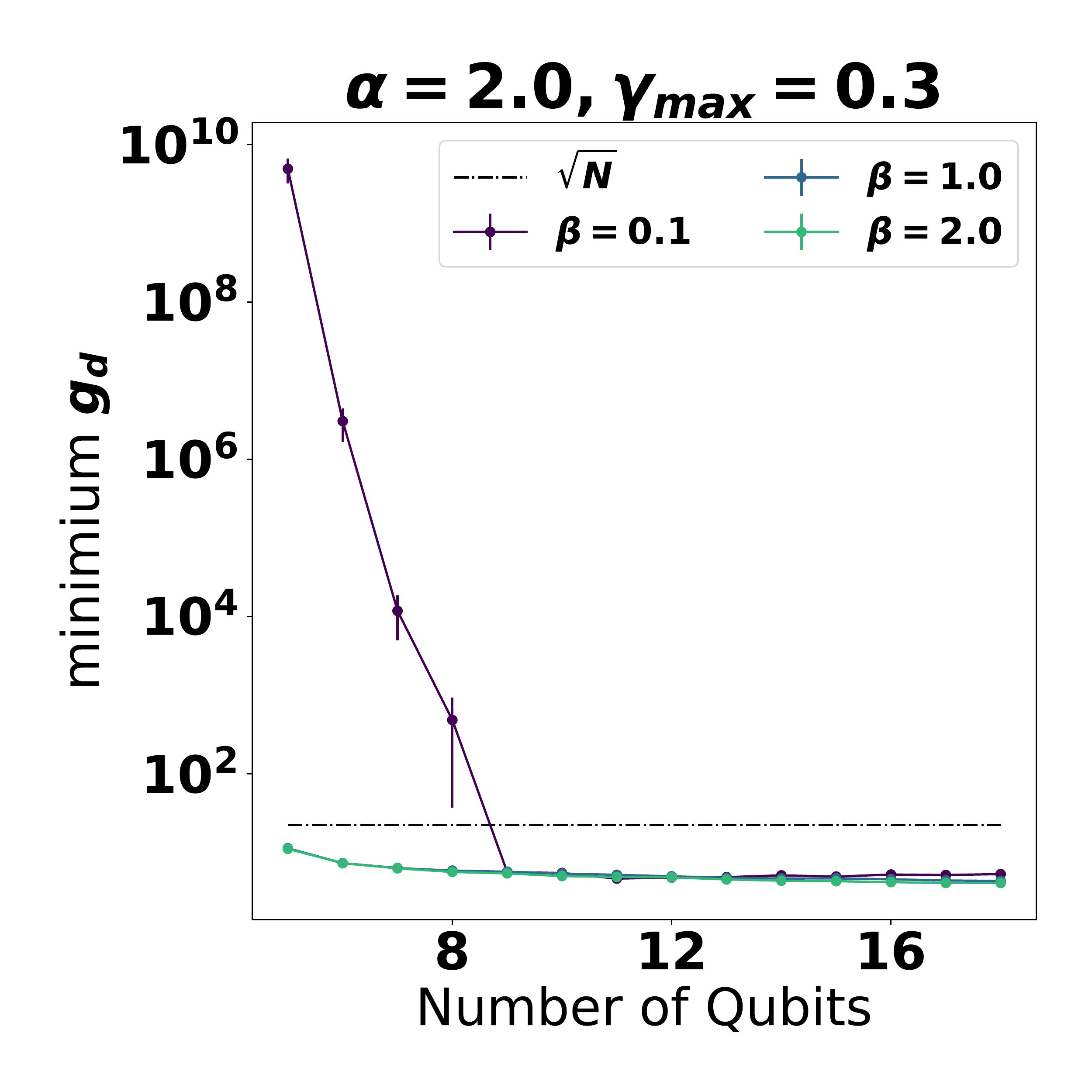}
\label{fig:vary_ab_gd_scaling-b}
}
\hspace{-20pt}
\subfloat[]{
\includegraphics[width=132pt]{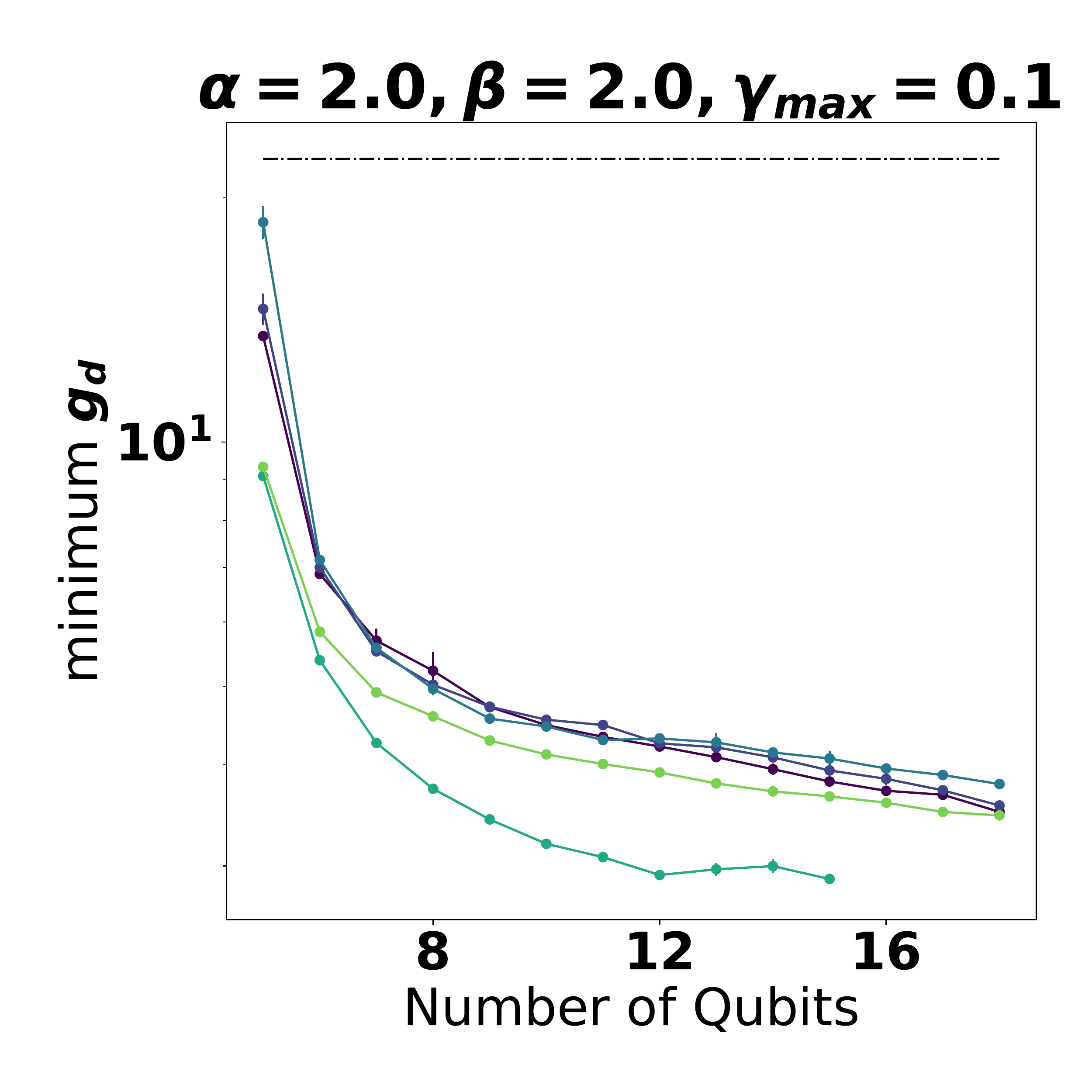}
\label{fig:vary_ab_gd_scaling-c}
}
\hspace{-20pt}
\subfloat[]{
\includegraphics[width=132pt]{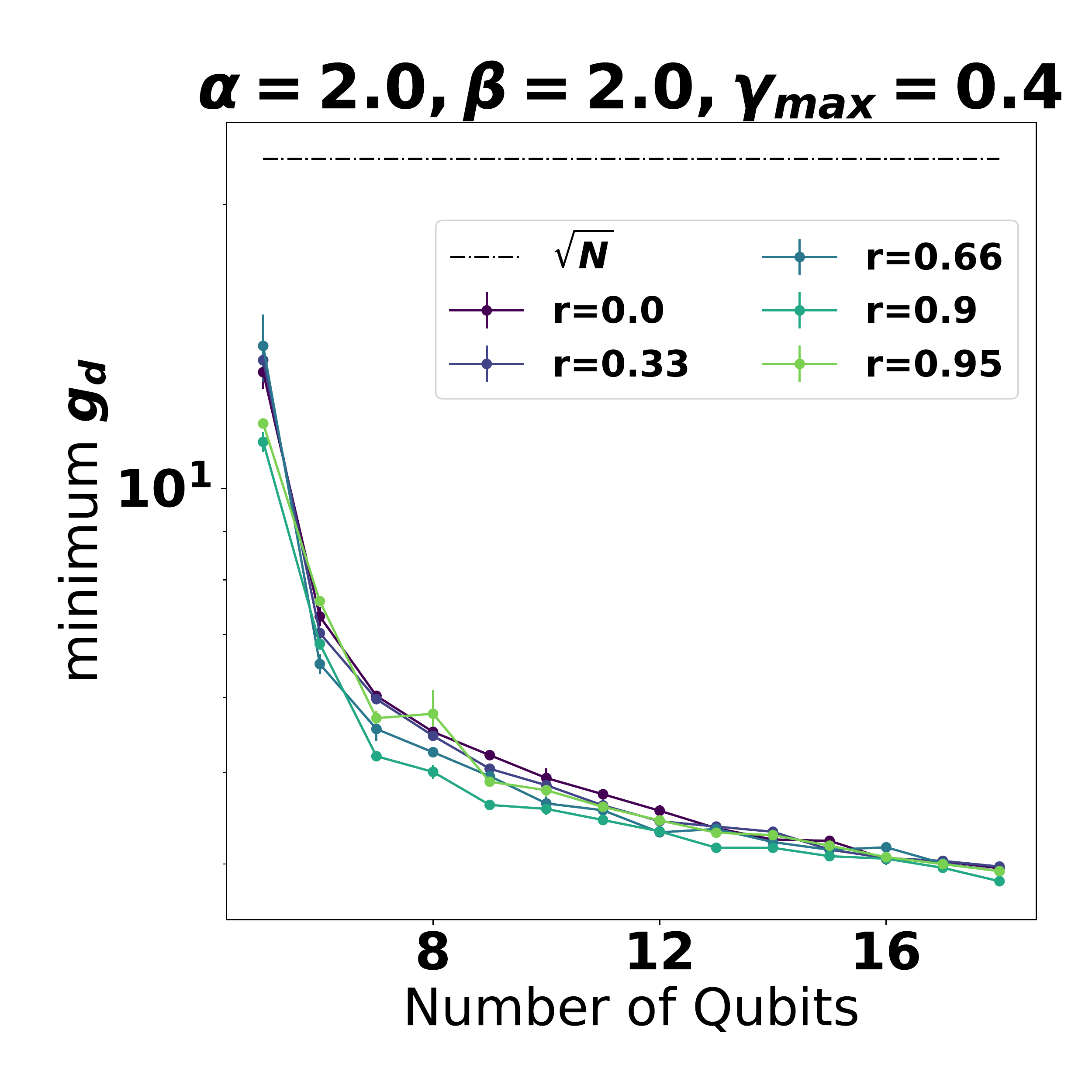}
\label{fig:vary_ab_gd_scaling-d}
}
\end{tabular}
\caption{Plots of the minimum geometric difference vs. number of qubits using the IQP style feature map. We show results for different data distributions and values of $\alpha$. In \protect\subref{fig:vary_ab_gd_scaling-a}  and \protect\subref{fig:vary_ab_gd_scaling-b}, we show results for uncorrelated data distributions while in \protect\subref{fig:vary_ab_gd_scaling-c}  and \protect\subref{fig:vary_ab_gd_scaling-d}  we show results for correlated data distributions. \protect\subref{fig:vary_ab_gd_scaling-a}  Minimum geometric difference, $g_d$, vs. number of qubits for $\beta=1.0$, $\gamma_{max}=0.3$ and varying $\alpha$. \protect\subref{fig:vary_ab_gd_scaling-b} Minimum geometric difference, $g_d$, vs. number of qubits for $\alpha=2.0$, $\gamma_{max}=0.3$ and varying $\beta$. Minimum geometric difference, $g_d$, vs. number of qubits for  $\alpha=2.0$, $\beta=2.0$ and \protect\subref{fig:vary_ab_gd_scaling-c} $\gamma_{max}=0.1$ (\protect\subref{fig:vary_ab_gd_scaling-d} $\gamma_{max}=0.4$) for varying correlation coefficient r.}
\label{fig:vary_ab_gd_scaling}
\end{figure*}

In Sec.~\ref{sec:fixed_alpha_beta}, we presented numerical results for the IQP style feature map with $\alpha=2.0$ and data distribution with $\beta=1.0$. In this section, we will expand our results to include IQP style feature maps with $\alpha \in \{0.5, 1.0, 2.0\}$ as well as uncorrelated data distributions with $\beta\in \{0.1, 1.0, 2.0\}$. For each $\alpha$ and $\beta$ pair, we repeat the analysis done in Fig.~\ref{fig:fixed_ab_curves} in order to extract the appropriate $\lambda$s to tune our IQP style feature map and keep $\gamma_{max}$ fixed. In Fig.~\ref{fig:vary_ab_gd_scaling}\subref{fig:vary_ab_gd_scaling-a}, We plot the minimum $g_d$ found vs. number of qubits for $\beta=1.0$, $\gamma_{max}=0.3$ and $\alpha\in \{0.5, 1.0, 2.0\}$. All three IQP style feature maps exhibit the same qualitative behavior where $g_d$ rapidly decays with increasing number of qubits to below the $\sqrt{N}$ line. In Fig.~\ref{fig:vary_ab_gd_scaling}\subref{fig:vary_ab_gd_scaling-b}, We plot the minimum $g_d$ found vs. number of qubits for $\alpha=2.0$, $\gamma_{max}=0.3$ and $\beta\in \{0.1, 1.0, 2.0\}$. Again, $g_d$ rapidly decays to below the $\sqrt{N}$ line with increasing number of qubits. While the higher kurtosis distribution with $\beta=0.1$ initially starts with much higher $g_d$, it rapidly decays to below $\sqrt{N}$. In both Fig.~\ref{fig:vary_ab_gd_scaling}\subref{fig:vary_ab_gd_scaling-a} and Fig.~\ref{fig:vary_ab_gd_scaling}\subref{fig:vary_ab_gd_scaling-b}, we observe a rapid decay in $g_d$ with increasing number of qubits to well below $\sqrt{N}$ indicating that the potential for a quantum advantage is diminishing with system size for this data point distribution and feature map combinations.

\subsubsection{\label{sec:correlated_data} $\alpha=2, \hspace{2pt} \beta=2$ correlated data}

In addition to uncorrelated data distributions, we present numerical results for correlated data distributions. We use data points sampled from the correlated distribution given by Eq.~\ref{correlated_gennorm} with $\beta=2.0$ and varying $r$. For our quantum kernel, we embed the data points in the IQP feature map with $\alpha=2.0$ while for our classical kernel we use the 'classical' IQP feature map with $\alpha=2.0$. In Fig~\ref{fig:vary_ab_gd_scaling}\subref{fig:vary_ab_gd_scaling-c}\subref{fig:vary_ab_gd_scaling-d}, we plot the minimum $g_d$ found between the quantum kernel and the classical kernel vs. number of qubits for $r \in \{0.0, 0.33, 0.66, 0.9, 0.95\}$ and for $\gamma_{max}=0.1$ and $\gamma_{max}=0.4$. For both $\gamma_{max}s$ and all $r$, the minimum $g_d$ starts below $\sqrt{N}$ and only decays further indicating that the possibility of an advantage for the quantum kernel decreases with an increasing number of qubits.

\subsection{\label{sec:heisenberg_experiments} Heisenberg Feature Map}

\begin{figure}
\centering

\includegraphics[width=238pt]{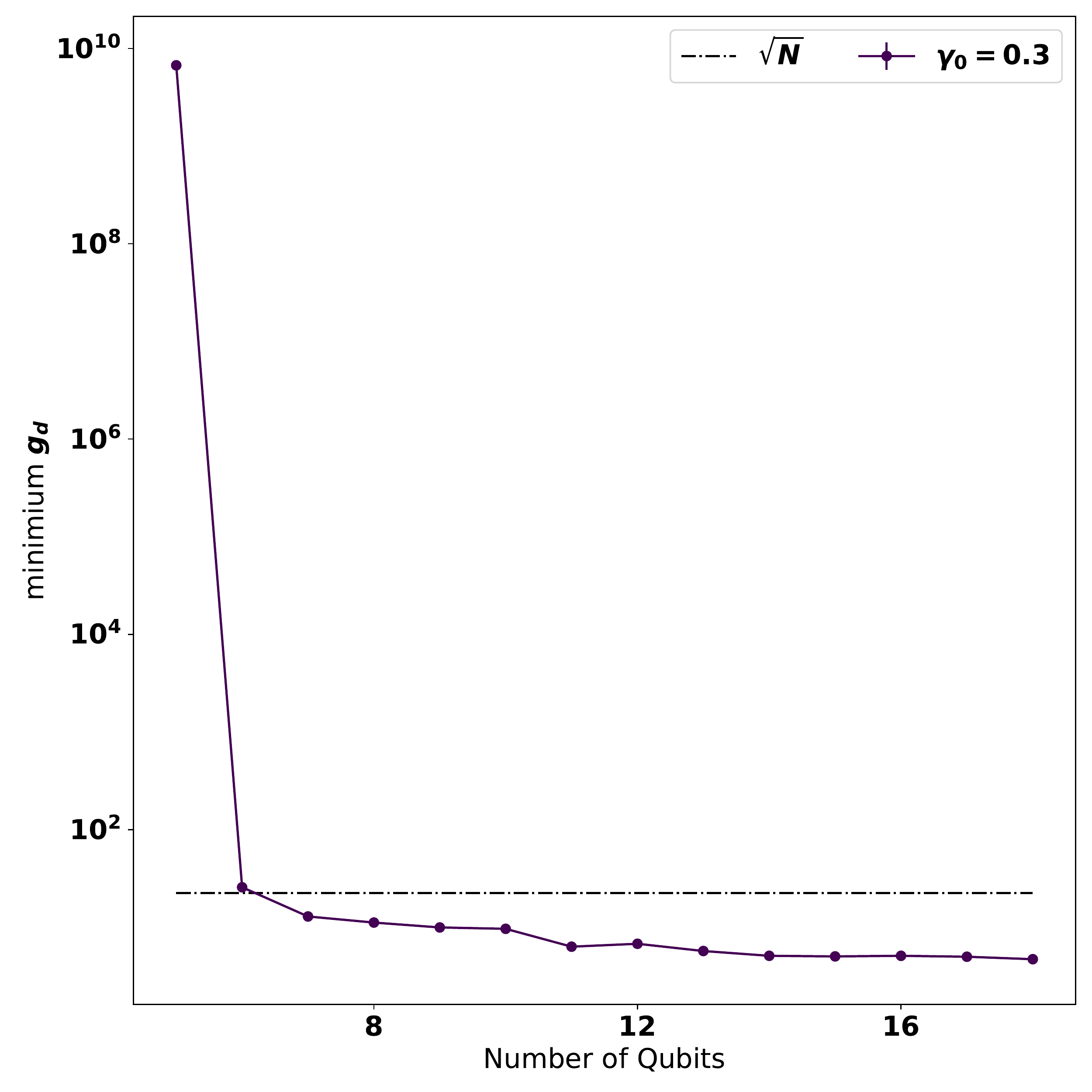}
\caption{Plot of the minimum $g_d$ vs. number of qubits for a the Heisenberg feature map (Eq.~\ref{hfeaturemap}) with $n_l=4$ and an uncorrelated data distribution with $\beta=1.0$.}
\label{fig:heisenberg_gd_scaling}
\end{figure}

In this section, we present numerical results for the Heisenberg feature map (Eq.~\ref{hfeaturemap}) and the classical fidelity kernel generated by the feature map in (Eq.~\ref{classical_hfeaturemap}). All kernel matrices use N=500 randomly sampled data points. We perform the same analysis as described in Sec.~\ref{sec:fixed_alpha_beta} and shown in Fig.~\ref{fig:fixed_ab_curves} for the Heisenberg feature map with $n_l=4$. In Fig.~\ref{fig:heisenberg_gd_scaling}, we plot the minimum $g_d$ vs. number of qubits for uncorrelated data distributions with $\beta=1.0$. We observe the same qualitative behavior as the IQP feature maps. The minimum $g_d$ found between quantum and the classical kernel decays rapidly with an increasing number of qubits to below $O(\sqrt{N})$. 

\section{Limitations}\label{sec:limitations}

The central observation we make is that numerically, the geometric difference between ``well-behaved'' quantum models and classical ones is small and goes down with the number of qubits. Here ``well-behaved'' is operationalized as ``able to generalize with a fixed number of training points''. There are three main potential limitations of the current study that we now discuss.

First, the condition that the number of training points (or, equivalently, $\gamma_{\max}$) stays constant with the number of qubits may be overly restrictive. Specifically, efficient generalization is possible as long as $\gamma_{\max}$ does not decay faster than polynomially with the number of qubits. Therefore, there may be a possibility of quantum advantage with kernels where $\gamma_{\max}$ decays polynomially and geometric difference stays large. However, this regime is difficult to probe due to only a small number of qubits accessible in simulation, which is why it is outside of the scope of the current study.

Second, we make a specific choice of technique to improve the generalization (namely, bandwidth tuning proposed in \cite{Shaydulin,Canatar}). While no other technique has been demonstrated to enable generalization of quantum kernel methods on a broad range of problems, it is possible that such technique will be proposed in the future. 

Third, we only consider a few of the best-performing quantum feature maps proposed in the literature. While we expect similar results to hold for any sufficiently expressive quantum feature maps, the decay of geometric difference need not hold in general for a carefully designed feature map. In particular, we do not expect to see a decay in geometric difference for the feature map based on Shor's algorithm that was proposed in \cite{Liu2021} for a learning version of the discrete logarithm problem. Whether or not such quantum feature maps exist for a broader range of machine learning problems remains an open question.

\section{\label{sec:discussion} Discussion}

In this work we provide numerical evidence that quantum (fidelity) kernels are unlikely to provide advantage over classical machine learning models. Unlike \cite{Huang2021}, we consider explicitly quantum kernels that perform well on classical data. Nonetheless our conclusion is similar to that of \cite{Huang2021}. Without further algorithmic advances, quantum advantage on classical data is unlikely.

The central challenge is need to encode inductive bias into the quantum model to prevent decay of the largest eigenvalue and ``flattening'' of the spectrum. While considering parameterized quantum circuits (also called ``Quantum Neural Networks'' or QNNs) may provide some benefit by using the training method to introduce inductive bias, in general it is not clear if it's easier to introduce structure into QNNs than into quantum feature maps. It may be possible to leverage group structure of the data~\cite{Larocca2022,2207.14413} to overcome this challenge, though applying it to a broad range of classical problems appears challenging. Without such bias, general-purpose techniques like bandwidth tuning~\cite{Shaydulin,Canatar} or projected kernels~\cite{Kubler2021,Huang2021} must be used, which do not appear to provide a path to quantum advantage on classical data.

\section*{Acknowledgements}

The authors thank Dylan Herman and other members of the Global Technology Applied Research center of JPMorgan Chase for helpful discussions.
This work was supported in part by the U.S.\ Department of Energy (DOE), Office of Science, Office of Advanced Scientific Computing Research AIDE-QC and FAR-QC projects and by the Argonne LDRD program under contract number DE-AC02-06CH11357. We gratefully acknowledge the computing resources provided on Bebop, a high-performance computing cluster operated by the Laboratory Computing Resource Center at Argonne National Laboratory.

\bibliographystyle{apsrev4-1}
\bibliography{main_pra}

\begin{thebibliography}{41}%
\makeatletter
\providecommand \@ifxundefined [1]{%
 \@ifx{#1\undefined}
}%
\providecommand \@ifnum [1]{%
 \ifnum #1\expandafter \@firstoftwo
 \else \expandafter \@secondoftwo
 \fi
}%
\providecommand \@ifx [1]{%
 \ifx #1\expandafter \@firstoftwo
 \else \expandafter \@secondoftwo
 \fi
}%
\providecommand \natexlab [1]{#1}%
\providecommand \enquote  [1]{``#1''}%
\providecommand \bibnamefont  [1]{#1}%
\providecommand \bibfnamefont [1]{#1}%
\providecommand \citenamefont [1]{#1}%
\providecommand \href@noop [0]{\@secondoftwo}%
\providecommand \href [0]{\begingroup \@sanitize@url \@href}%
\providecommand \@href[1]{\@@startlink{#1}\@@href}%
\providecommand \@@href[1]{\endgroup#1\@@endlink}%
\providecommand \@sanitize@url [0]{\catcode `\\12\catcode `\$12\catcode
  `\&12\catcode `\#12\catcode `\^12\catcode `\_12\catcode `\%12\relax}%
\providecommand \@@startlink[1]{}%
\providecommand \@@endlink[0]{}%
\providecommand \url  [0]{\begingroup\@sanitize@url \@url }%
\providecommand \@url [1]{\endgroup\@href {#1}{\urlprefix }}%
\providecommand \urlprefix  [0]{URL }%
\providecommand \Eprint [0]{\href }%
\providecommand \doibase [0]{http://dx.doi.org/}%
\providecommand \selectlanguage [0]{\@gobble}%
\providecommand \bibinfo  [0]{\@secondoftwo}%
\providecommand \bibfield  [0]{\@secondoftwo}%
\providecommand \translation [1]{[#1]}%
\providecommand \BibitemOpen [0]{}%
\providecommand \bibitemStop [0]{}%
\providecommand \bibitemNoStop [0]{.\EOS\space}%
\providecommand \EOS [0]{\spacefactor3000\relax}%
\providecommand \BibitemShut  [1]{\csname bibitem#1\endcsname}%
\let\auto@bib@innerbib\@empty
\bibitem [{\citenamefont {Nielsen}\ and\ \citenamefont
  {Chuang}(2011)}]{nielsen2011quantum}%
  \BibitemOpen
  \bibfield  {author} {\bibinfo {author} {\bibfnamefont {M.~A.}\ \bibnamefont
  {Nielsen}}\ and\ \bibinfo {author} {\bibfnamefont {I.~L.}\ \bibnamefont
  {Chuang}},\ }\href {\doibase 10.1017/CBO9780511976667} {\emph {\bibinfo
  {title} {Quantum Computation and Quantum Information}}}\ (\bibinfo
  {publisher} {Cambridge University Press},\ \bibinfo {year}
  {2011})\BibitemShut {NoStop}%
\bibitem [{\citenamefont {Biamonte}\ \emph {et~al.}(2017)\citenamefont
  {Biamonte}, \citenamefont {Wittek}, \citenamefont {Pancotti}, \citenamefont
  {Rebentrost}, \citenamefont {Wiebe},\ and\ \citenamefont
  {Lloyd}}]{Biamonte2017}%
  \BibitemOpen
  \bibfield  {author} {\bibinfo {author} {\bibfnamefont {J.}~\bibnamefont
  {Biamonte}}, \bibinfo {author} {\bibfnamefont {P.}~\bibnamefont {Wittek}},
  \bibinfo {author} {\bibfnamefont {N.}~\bibnamefont {Pancotti}}, \bibinfo
  {author} {\bibfnamefont {P.}~\bibnamefont {Rebentrost}}, \bibinfo {author}
  {\bibfnamefont {N.}~\bibnamefont {Wiebe}}, \ and\ \bibinfo {author}
  {\bibfnamefont {S.}~\bibnamefont {Lloyd}},\ }\href {\doibase
  10.1038/nature23474} {\bibfield  {journal} {\bibinfo  {journal} {Nature}\
  }\textbf {\bibinfo {volume} {549}},\ \bibinfo {pages} {195} (\bibinfo {year}
  {2017})}\BibitemShut {NoStop}%
\bibitem [{\citenamefont {Pistoia}\ \emph {et~al.}(2021)\citenamefont
  {Pistoia}, \citenamefont {Ahmad}, \citenamefont {Ajagekar}, \citenamefont
  {Buts}, \citenamefont {Chakrabarti}, \citenamefont {Herman}, \citenamefont
  {Hu}, \citenamefont {Jena}, \citenamefont {Minssen}, \citenamefont {Niroula},
  \citenamefont {Rattew}, \citenamefont {Sun},\ and\ \citenamefont
  {Yalovetzky}}]{Pistoia2021}%
  \BibitemOpen
  \bibfield  {author} {\bibinfo {author} {\bibfnamefont {M.}~\bibnamefont
  {Pistoia}}, \bibinfo {author} {\bibfnamefont {S.~F.}\ \bibnamefont {Ahmad}},
  \bibinfo {author} {\bibfnamefont {A.}~\bibnamefont {Ajagekar}}, \bibinfo
  {author} {\bibfnamefont {A.}~\bibnamefont {Buts}}, \bibinfo {author}
  {\bibfnamefont {S.}~\bibnamefont {Chakrabarti}}, \bibinfo {author}
  {\bibfnamefont {D.}~\bibnamefont {Herman}}, \bibinfo {author} {\bibfnamefont
  {S.}~\bibnamefont {Hu}}, \bibinfo {author} {\bibfnamefont {A.}~\bibnamefont
  {Jena}}, \bibinfo {author} {\bibfnamefont {P.}~\bibnamefont {Minssen}},
  \bibinfo {author} {\bibfnamefont {P.}~\bibnamefont {Niroula}}, \bibinfo
  {author} {\bibfnamefont {A.}~\bibnamefont {Rattew}}, \bibinfo {author}
  {\bibfnamefont {Y.}~\bibnamefont {Sun}}, \ and\ \bibinfo {author}
  {\bibfnamefont {R.}~\bibnamefont {Yalovetzky}},\ }in\ \href {\doibase
  10.1109/iccad51958.2021.9643469} {\emph {\bibinfo {booktitle} {2021
  {IEEE}/{ACM} International Conference On Computer Aided Design ({ICCAD})}}}\
  (\bibinfo  {publisher} {{IEEE}},\ \bibinfo {year} {2021})\BibitemShut
  {NoStop}%
\bibitem [{\citenamefont {Herman}\ \emph {et~al.}(2022)\citenamefont {Herman},
  \citenamefont {Googin}, \citenamefont {Liu}, \citenamefont {Galda},
  \citenamefont {Safro}, \citenamefont {Sun}, \citenamefont {Pistoia},\ and\
  \citenamefont {Alexeev}}]{2201.02773}%
  \BibitemOpen
  \bibfield  {author} {\bibinfo {author} {\bibfnamefont {D.}~\bibnamefont
  {Herman}}, \bibinfo {author} {\bibfnamefont {C.}~\bibnamefont {Googin}},
  \bibinfo {author} {\bibfnamefont {X.}~\bibnamefont {Liu}}, \bibinfo {author}
  {\bibfnamefont {A.}~\bibnamefont {Galda}}, \bibinfo {author} {\bibfnamefont
  {I.}~\bibnamefont {Safro}}, \bibinfo {author} {\bibfnamefont
  {Y.}~\bibnamefont {Sun}}, \bibinfo {author} {\bibfnamefont {M.}~\bibnamefont
  {Pistoia}}, \ and\ \bibinfo {author} {\bibfnamefont {Y.}~\bibnamefont
  {Alexeev}},\ }\href@noop {} {\bibfield  {journal} {\bibinfo  {journal}
  {arXiv:2201.02773}\ } (\bibinfo {year} {2022})}\BibitemShut {NoStop}%
\bibitem [{\citenamefont {Huang}\ \emph {et~al.}(2022)\citenamefont {Huang},
  \citenamefont {Broughton}, \citenamefont {Cotler}, \citenamefont {Chen},
  \citenamefont {Li}, \citenamefont {Mohseni}, \citenamefont {Neven},
  \citenamefont {Babbush}, \citenamefont {Kueng}, \citenamefont {Preskill},\
  and\ \citenamefont {McClean}}]{Huang2022}%
  \BibitemOpen
  \bibfield  {author} {\bibinfo {author} {\bibfnamefont {H.~Y.}\ \bibnamefont
  {Huang}}, \bibinfo {author} {\bibfnamefont {M.}~\bibnamefont {Broughton}},
  \bibinfo {author} {\bibfnamefont {J.}~\bibnamefont {Cotler}}, \bibinfo
  {author} {\bibfnamefont {S.}~\bibnamefont {Chen}}, \bibinfo {author}
  {\bibfnamefont {J.}~\bibnamefont {Li}}, \bibinfo {author} {\bibfnamefont
  {M.}~\bibnamefont {Mohseni}}, \bibinfo {author} {\bibfnamefont
  {H.}~\bibnamefont {Neven}}, \bibinfo {author} {\bibfnamefont
  {R.}~\bibnamefont {Babbush}}, \bibinfo {author} {\bibfnamefont
  {R.}~\bibnamefont {Kueng}}, \bibinfo {author} {\bibfnamefont
  {J.}~\bibnamefont {Preskill}}, \ and\ \bibinfo {author} {\bibfnamefont
  {J.~R.}\ \bibnamefont {McClean}},\ }\href {\doibase
  10.1126/SCIENCE.ABN7293/SUPPL_FILE/SCIENCE.ABN7293_SM.PDF} {\bibfield
  {journal} {\bibinfo  {journal} {Science}\ }\textbf {\bibinfo {volume}
  {376}},\ \bibinfo {pages} {1182} (\bibinfo {year} {2022})},\ \Eprint
  {http://arxiv.org/abs/2112.00778} {arXiv:2112.00778} \BibitemShut {NoStop}%
\bibitem [{\citenamefont {K{\"{u}}bler}\ \emph {et~al.}(2021)\citenamefont
  {K{\"{u}}bler}, \citenamefont {Buchholz},\ and\ \citenamefont
  {Sch{\"{o}}lkopf}}]{Kubler2021}%
  \BibitemOpen
  \bibfield  {author} {\bibinfo {author} {\bibfnamefont {J.~M.}\ \bibnamefont
  {K{\"{u}}bler}}, \bibinfo {author} {\bibfnamefont {S.}~\bibnamefont
  {Buchholz}}, \ and\ \bibinfo {author} {\bibfnamefont {B.}~\bibnamefont
  {Sch{\"{o}}lkopf}},\ }\href@noop {} {\  (\bibinfo {year} {2021})},\ \Eprint
  {http://arxiv.org/abs/2106.03747v2} {arXiv:2106.03747v2} \BibitemShut
  {NoStop}%
\bibitem [{\citenamefont {Liu}\ \emph {et~al.}(2021)\citenamefont {Liu},
  \citenamefont {Arunachalam},\ and\ \citenamefont {Temme}}]{Liu2021}%
  \BibitemOpen
  \bibfield  {author} {\bibinfo {author} {\bibfnamefont {Y.}~\bibnamefont
  {Liu}}, \bibinfo {author} {\bibfnamefont {S.}~\bibnamefont {Arunachalam}}, \
  and\ \bibinfo {author} {\bibfnamefont {K.}~\bibnamefont {Temme}},\ }\href
  {\doibase 10.1038/s41567-021-01287-z} {\bibfield  {journal} {\bibinfo
  {journal} {Nature Physics 2021 17:9}\ }\textbf {\bibinfo {volume} {17}},\
  \bibinfo {pages} {1013} (\bibinfo {year} {2021})},\ \Eprint
  {http://arxiv.org/abs/2010.02174} {arXiv:2010.02174} \BibitemShut {NoStop}%
\bibitem [{\citenamefont {Schmidhuber}\ and\ \citenamefont
  {Lloyd}(2022)}]{Lloyd2022TDA}%
  \BibitemOpen
  \bibfield  {author} {\bibinfo {author} {\bibfnamefont {A.}~\bibnamefont
  {Schmidhuber}}\ and\ \bibinfo {author} {\bibfnamefont {S.}~\bibnamefont
  {Lloyd}},\ }\href@noop {} {\bibfield  {journal} {\bibinfo  {journal}
  {arXiv:2209.14286}\ } (\bibinfo {year} {2022})}\BibitemShut {NoStop}%
\bibitem [{\citenamefont {Schuld}\ and\ \citenamefont
  {Killoran}(2019)}]{Schuld2019}%
  \BibitemOpen
  \bibfield  {author} {\bibinfo {author} {\bibfnamefont {M.}~\bibnamefont
  {Schuld}}\ and\ \bibinfo {author} {\bibfnamefont {N.}~\bibnamefont
  {Killoran}},\ }\href {\doibase 10.1103/physrevlett.122.040504} {\bibfield
  {journal} {\bibinfo  {journal} {Phys. Rev. Lett.}\ }\textbf {\bibinfo
  {volume} {122}},\ \bibinfo {pages} {040504} (\bibinfo {year} {2019})},\
  \Eprint {http://arxiv.org/abs/1803.07128} {arXiv:1803.07128} \BibitemShut
  {NoStop}%
\bibitem [{\citenamefont {Park}\ \emph {et~al.}(2020)\citenamefont {Park},
  \citenamefont {Blank},\ and\ \citenamefont {Petruccione}}]{Park2020}%
  \BibitemOpen
  \bibfield  {author} {\bibinfo {author} {\bibfnamefont {D.~K.}\ \bibnamefont
  {Park}}, \bibinfo {author} {\bibfnamefont {C.}~\bibnamefont {Blank}}, \ and\
  \bibinfo {author} {\bibfnamefont {F.}~\bibnamefont {Petruccione}},\ }\href
  {\doibase 10.1016/J.PHYSLETA.2020.126422} {\bibfield  {journal} {\bibinfo
  {journal} {Physics Letters A}\ }\textbf {\bibinfo {volume} {384}},\ \bibinfo
  {pages} {126422} (\bibinfo {year} {2020})},\ \Eprint
  {http://arxiv.org/abs/2004.03489} {arXiv:2004.03489} \BibitemShut {NoStop}%
\bibitem [{\citenamefont {Havl{\'{i}}{\v{c}}ek}\ \emph
  {et~al.}(2019)\citenamefont {Havl{\'{i}}{\v{c}}ek}, \citenamefont
  {C{\'{o}}rcoles}, \citenamefont {Temme}, \citenamefont {Harrow},
  \citenamefont {Kandala}, \citenamefont {Chow},\ and\ \citenamefont
  {Gambetta}}]{Havlicek2019}%
  \BibitemOpen
  \bibfield  {author} {\bibinfo {author} {\bibfnamefont {V.}~\bibnamefont
  {Havl{\'{i}}{\v{c}}ek}}, \bibinfo {author} {\bibfnamefont {A.~D.}\
  \bibnamefont {C{\'{o}}rcoles}}, \bibinfo {author} {\bibfnamefont
  {K.}~\bibnamefont {Temme}}, \bibinfo {author} {\bibfnamefont {A.~W.}\
  \bibnamefont {Harrow}}, \bibinfo {author} {\bibfnamefont {A.}~\bibnamefont
  {Kandala}}, \bibinfo {author} {\bibfnamefont {J.~M.}\ \bibnamefont {Chow}}, \
  and\ \bibinfo {author} {\bibfnamefont {J.~M.}\ \bibnamefont {Gambetta}},\
  }\href {\doibase 10.1038/s41586-019-0980-2} {\bibfield  {journal} {\bibinfo
  {journal} {Nature 2019 567:7747}\ }\textbf {\bibinfo {volume} {567}},\
  \bibinfo {pages} {209} (\bibinfo {year} {2019})},\ \Eprint
  {http://arxiv.org/abs/1804.11326} {arXiv:1804.11326} \BibitemShut {NoStop}%
\bibitem [{\citenamefont {Raghu}\ \emph {et~al.}(2017)\citenamefont {Raghu},
  \citenamefont {Poole}, \citenamefont {Kleinberg}, \citenamefont {Ganguli},\
  and\ \citenamefont {Dickstein}}]{Raghu2017}%
  \BibitemOpen
  \bibfield  {author} {\bibinfo {author} {\bibfnamefont {M.}~\bibnamefont
  {Raghu}}, \bibinfo {author} {\bibfnamefont {B.}~\bibnamefont {Poole}},
  \bibinfo {author} {\bibfnamefont {J.}~\bibnamefont {Kleinberg}}, \bibinfo
  {author} {\bibfnamefont {S.}~\bibnamefont {Ganguli}}, \ and\ \bibinfo
  {author} {\bibfnamefont {J.~S.}\ \bibnamefont {Dickstein}},\ }\href@noop {}
  {\emph {\bibinfo {title} {{On the Expressive Power of Deep Neural
  Networks}}}},\ \bibinfo {type} {Tech. Rep.}\ (\bibinfo {year} {2017})\
  \Eprint {http://arxiv.org/abs/1606.05336v6} {arXiv:1606.05336v6} \BibitemShut
  {NoStop}%
\bibitem [{\citenamefont {Schuld}\ \emph {et~al.}(2021)\citenamefont {Schuld},
  \citenamefont {Sweke},\ and\ \citenamefont {Meyer}}]{Schuld2021}%
  \BibitemOpen
  \bibfield  {author} {\bibinfo {author} {\bibfnamefont {M.}~\bibnamefont
  {Schuld}}, \bibinfo {author} {\bibfnamefont {R.}~\bibnamefont {Sweke}}, \
  and\ \bibinfo {author} {\bibfnamefont {J.~J.}\ \bibnamefont {Meyer}},\
  }\href@noop {} {\  (\bibinfo {year} {2021})},\ \Eprint
  {http://arxiv.org/abs/2008.08605v2} {arXiv:2008.08605v2} \BibitemShut
  {NoStop}%
\bibitem [{\citenamefont {Blank}\ \emph {et~al.}(2020)\citenamefont {Blank},
  \citenamefont {Park}, \citenamefont {Rhee},\ and\ \citenamefont
  {Petruccione}}]{Blank2020}%
  \BibitemOpen
  \bibfield  {author} {\bibinfo {author} {\bibfnamefont {C.}~\bibnamefont
  {Blank}}, \bibinfo {author} {\bibfnamefont {D.~K.}\ \bibnamefont {Park}},
  \bibinfo {author} {\bibfnamefont {J.~K.~K.}\ \bibnamefont {Rhee}}, \ and\
  \bibinfo {author} {\bibfnamefont {F.}~\bibnamefont {Petruccione}},\ }\href
  {\doibase 10.1038/s41534-020-0272-6} {\bibfield  {journal} {\bibinfo
  {journal} {npj Quantum Information 2020 6:1}\ }\textbf {\bibinfo {volume}
  {6}},\ \bibinfo {pages} {1} (\bibinfo {year} {2020})},\ \Eprint
  {http://arxiv.org/abs/1909.02611} {arXiv:1909.02611} \BibitemShut {NoStop}%
\bibitem [{\citenamefont {Abbas}\ \emph {et~al.}(2020)\citenamefont {Abbas},
  \citenamefont {Sutter}, \citenamefont {Zoufal}, \citenamefont {Lucchi},
  \citenamefont {Figalli},\ and\ \citenamefont {Woerner}}]{Abbas2020}%
  \BibitemOpen
  \bibfield  {author} {\bibinfo {author} {\bibfnamefont {A.}~\bibnamefont
  {Abbas}}, \bibinfo {author} {\bibfnamefont {D.}~\bibnamefont {Sutter}},
  \bibinfo {author} {\bibfnamefont {C.}~\bibnamefont {Zoufal}}, \bibinfo
  {author} {\bibfnamefont {A.}~\bibnamefont {Lucchi}}, \bibinfo {author}
  {\bibfnamefont {A.}~\bibnamefont {Figalli}}, \ and\ \bibinfo {author}
  {\bibfnamefont {S.}~\bibnamefont {Woerner}},\ }\href@noop {} {\  (\bibinfo
  {year} {2020})},\ \Eprint {http://arxiv.org/abs/2011.00027v1}
  {arXiv:2011.00027v1} \BibitemShut {NoStop}%
\bibitem [{\citenamefont {Banchi}\ \emph {et~al.}(2021)\citenamefont {Banchi},
  \citenamefont {Pereira},\ and\ \citenamefont {Pirandola}}]{Banchi2021}%
  \BibitemOpen
  \bibfield  {author} {\bibinfo {author} {\bibfnamefont {L.}~\bibnamefont
  {Banchi}}, \bibinfo {author} {\bibfnamefont {J.}~\bibnamefont {Pereira}}, \
  and\ \bibinfo {author} {\bibfnamefont {S.}~\bibnamefont {Pirandola}},\ }\href
  {\doibase 10.1103/PRXQuantum.2.040321} {\bibfield  {journal} {\bibinfo
  {journal} {PRX QUANTUM}\ }\textbf {\bibinfo {volume} {2}},\ \bibinfo {pages}
  {40321} (\bibinfo {year} {2021})}\BibitemShut {NoStop}%
\bibitem [{\citenamefont {Canatar}\ \emph {et~al.}()\citenamefont {Canatar},
  \citenamefont {Peters}, \citenamefont {Pehlevan}, \citenamefont {Wild},\ and\
  \citenamefont {Shaydulin}}]{Canatar}%
  \BibitemOpen
  \bibfield  {author} {\bibinfo {author} {\bibfnamefont {A.}~\bibnamefont
  {Canatar}}, \bibinfo {author} {\bibfnamefont {E.}~\bibnamefont {Peters}},
  \bibinfo {author} {\bibfnamefont {C.}~\bibnamefont {Pehlevan}}, \bibinfo
  {author} {\bibfnamefont {S.~M.}\ \bibnamefont {Wild}}, \ and\ \bibinfo
  {author} {\bibfnamefont {R.}~\bibnamefont {Shaydulin}},\ }\href@noop {} {\
  }\Eprint {http://arxiv.org/abs/2206.06686v2} {arXiv:2206.06686v2}
  \BibitemShut {NoStop}%
\bibitem [{\citenamefont {Huang}\ \emph {et~al.}(2021)\citenamefont {Huang},
  \citenamefont {Broughton}, \citenamefont {Mohseni}, \citenamefont {Babbush},
  \citenamefont {Boixo}, \citenamefont {Neven},\ and\ \citenamefont
  {Mcclean}}]{Huang2021}%
  \BibitemOpen
  \bibfield  {author} {\bibinfo {author} {\bibfnamefont {H.-Y.}\ \bibnamefont
  {Huang}}, \bibinfo {author} {\bibfnamefont {M.}~\bibnamefont {Broughton}},
  \bibinfo {author} {\bibfnamefont {M.}~\bibnamefont {Mohseni}}, \bibinfo
  {author} {\bibfnamefont {R.}~\bibnamefont {Babbush}}, \bibinfo {author}
  {\bibfnamefont {S.}~\bibnamefont {Boixo}}, \bibinfo {author} {\bibfnamefont
  {H.}~\bibnamefont {Neven}}, \ and\ \bibinfo {author} {\bibfnamefont {J.~R.}\
  \bibnamefont {Mcclean}},\ }\href {\doibase 10.1038/s41467-021-22539-9}
  {\bibfield  {journal} {\bibinfo  {journal} {Nature}\ } (\bibinfo {year}
  {2021}),\ 10.1038/s41467-021-22539-9}\BibitemShut {NoStop}%
\bibitem [{\citenamefont {Paine}\ \emph {et~al.}(2022)\citenamefont {Paine},
  \citenamefont {Elfving},\ and\ \citenamefont {Kyriienko}}]{Paine2022}%
  \BibitemOpen
  \bibfield  {author} {\bibinfo {author} {\bibfnamefont {A.~E.}\ \bibnamefont
  {Paine}}, \bibinfo {author} {\bibfnamefont {V.~E.}\ \bibnamefont {Elfving}},
  \ and\ \bibinfo {author} {\bibfnamefont {O.}~\bibnamefont {Kyriienko}},\
  }\href@noop {} {\  (\bibinfo {year} {2022})},\ \Eprint
  {http://arxiv.org/abs/2203.08884v1} {arXiv:2203.08884v1} \BibitemShut
  {NoStop}%
\bibitem [{\citenamefont {Peters}\ and\ \citenamefont
  {Schuld}(2022)}]{Peters2022}%
  \BibitemOpen
  \bibfield  {author} {\bibinfo {author} {\bibfnamefont {E.}~\bibnamefont
  {Peters}}\ and\ \bibinfo {author} {\bibfnamefont {M.}~\bibnamefont
  {Schuld}},\ }\href@noop {} {\  (\bibinfo {year} {2022})},\ \Eprint
  {http://arxiv.org/abs/2209.05523v1} {arXiv:2209.05523v1} \BibitemShut
  {NoStop}%
\bibitem [{\citenamefont {Shaydulin}\ and\ \citenamefont {Wild}()}]{Shaydulin}%
  \BibitemOpen
  \bibfield  {author} {\bibinfo {author} {\bibfnamefont {R.}~\bibnamefont
  {Shaydulin}}\ and\ \bibinfo {author} {\bibfnamefont {S.~M.}\ \bibnamefont
  {Wild}},\ }\href@noop {} {\ }\Eprint {http://arxiv.org/abs/2111.05451v3}
  {arXiv:2111.05451v3} \BibitemShut {NoStop}%
\bibitem [{\citenamefont {Shirai}\ \emph {et~al.}()\citenamefont {Shirai},
  \citenamefont {Kubo}, \citenamefont {Mitarai},\ and\ \citenamefont
  {Fujii}}]{Shirai}%
  \BibitemOpen
  \bibfield  {author} {\bibinfo {author} {\bibfnamefont {N.}~\bibnamefont
  {Shirai}}, \bibinfo {author} {\bibfnamefont {K.}~\bibnamefont {Kubo}},
  \bibinfo {author} {\bibfnamefont {K.}~\bibnamefont {Mitarai}}, \ and\
  \bibinfo {author} {\bibfnamefont {K.}~\bibnamefont {Fujii}},\ }\href@noop {}
  {\ }\Eprint {http://arxiv.org/abs/2111.02951v1} {arXiv:2111.02951v1}
  \BibitemShut {NoStop}%
\bibitem [{\citenamefont {Huang}\ \emph {et~al.}(2020)\citenamefont {Huang},
  \citenamefont {Kueng},\ and\ \citenamefont {Preskill}}]{Huang2020}%
  \BibitemOpen
  \bibfield  {author} {\bibinfo {author} {\bibfnamefont {H.~Y.}\ \bibnamefont
  {Huang}}, \bibinfo {author} {\bibfnamefont {R.}~\bibnamefont {Kueng}}, \ and\
  \bibinfo {author} {\bibfnamefont {J.}~\bibnamefont {Preskill}},\ }\href
  {\doibase 10.1038/s41567-020-0932-7} {\bibfield  {journal} {\bibinfo
  {journal} {Nature Physics 2020 16:10}\ }\textbf {\bibinfo {volume} {16}},\
  \bibinfo {pages} {1050} (\bibinfo {year} {2020})},\ \Eprint
  {http://arxiv.org/abs/2002.08953} {arXiv:2002.08953} \BibitemShut {NoStop}%
\bibitem [{\citenamefont {Canatar}\ \emph {et~al.}(2021)\citenamefont
  {Canatar}, \citenamefont {Bordelon},\ and\ \citenamefont
  {Pehlevan}}]{canatar2021spectral}%
  \BibitemOpen
  \bibfield  {author} {\bibinfo {author} {\bibfnamefont {A.}~\bibnamefont
  {Canatar}}, \bibinfo {author} {\bibfnamefont {B.}~\bibnamefont {Bordelon}}, \
  and\ \bibinfo {author} {\bibfnamefont {C.}~\bibnamefont {Pehlevan}},\ }\href
  {\doibase 10.1038/s41467-021-23103-1} {\bibfield  {journal} {\bibinfo
  {journal} {Nature Communications}\ }\textbf {\bibinfo {volume} {12}},\
  \bibinfo {pages} {1} (\bibinfo {year} {2021})}\BibitemShut {NoStop}%
\bibitem [{\citenamefont {Cerezo}\ \emph {et~al.}(2022)\citenamefont {Cerezo},
  \citenamefont {Verdon}, \citenamefont {Huang}, \citenamefont {Cincio},\ and\
  \citenamefont {Coles}}]{Cerezo2022}%
  \BibitemOpen
  \bibfield  {author} {\bibinfo {author} {\bibfnamefont {M.}~\bibnamefont
  {Cerezo}}, \bibinfo {author} {\bibfnamefont {G.}~\bibnamefont {Verdon}},
  \bibinfo {author} {\bibfnamefont {H.-Y.}\ \bibnamefont {Huang}}, \bibinfo
  {author} {\bibfnamefont {L.}~\bibnamefont {Cincio}}, \ and\ \bibinfo {author}
  {\bibfnamefont {P.~J.}\ \bibnamefont {Coles}},\ }\href {\doibase
  10.1038/s43588-022-00311-3} {\bibfield  {journal} {\bibinfo  {journal}
  {Nature Computational Science 2022 2:9}\ }\textbf {\bibinfo {volume} {2}},\
  \bibinfo {pages} {567} (\bibinfo {year} {2022})}\BibitemShut {NoStop}%
\bibitem [{\citenamefont {Cerezo}\ \emph {et~al.}(2021)\citenamefont {Cerezo},
  \citenamefont {Sone}, \citenamefont {Volkoff}, \citenamefont {Cincio},\ and\
  \citenamefont {Coles}}]{Cerezo2021a}%
  \BibitemOpen
  \bibfield  {author} {\bibinfo {author} {\bibfnamefont {M.}~\bibnamefont
  {Cerezo}}, \bibinfo {author} {\bibfnamefont {A.}~\bibnamefont {Sone}},
  \bibinfo {author} {\bibfnamefont {T.}~\bibnamefont {Volkoff}}, \bibinfo
  {author} {\bibfnamefont {L.}~\bibnamefont {Cincio}}, \ and\ \bibinfo {author}
  {\bibfnamefont {P.~J.}\ \bibnamefont {Coles}},\ }\href {\doibase
  10.1038/s41467-021-21728-w} {\bibfield  {journal} {\bibinfo  {journal}
  {Nature Communications 2021 12:1}\ }\textbf {\bibinfo {volume} {12}},\
  \bibinfo {pages} {1} (\bibinfo {year} {2021})},\ \Eprint
  {http://arxiv.org/abs/2001.00550} {arXiv:2001.00550} \BibitemShut {NoStop}%
\bibitem [{\citenamefont {Cerezo}\ and\ \citenamefont
  {Coles}(2021)}]{Cerezo2021}%
  \BibitemOpen
  \bibfield  {author} {\bibinfo {author} {\bibfnamefont {M.}~\bibnamefont
  {Cerezo}}\ and\ \bibinfo {author} {\bibfnamefont {P.~J.}\ \bibnamefont
  {Coles}},\ }\href {\doibase 10.1088/2058-9565/ABF51A} {\bibfield  {journal}
  {\bibinfo  {journal} {Quantum Science and Technology}\ }\textbf {\bibinfo
  {volume} {6}},\ \bibinfo {pages} {035006} (\bibinfo {year} {2021})},\ \Eprint
  {http://arxiv.org/abs/2008.07454} {arXiv:2008.07454} \BibitemShut {NoStop}%
\bibitem [{\citenamefont {Caro}\ \emph {et~al.}(2022)\citenamefont {Caro},
  \citenamefont {Huang}, \citenamefont {Cerezo}, \citenamefont {Sharma},
  \citenamefont {Sornborger}, \citenamefont {Cincio},\ and\ \citenamefont
  {Coles}}]{Caro2022}%
  \BibitemOpen
  \bibfield  {author} {\bibinfo {author} {\bibfnamefont {M.~C.}\ \bibnamefont
  {Caro}}, \bibinfo {author} {\bibfnamefont {H.-Y.}\ \bibnamefont {Huang}},
  \bibinfo {author} {\bibfnamefont {M.}~\bibnamefont {Cerezo}}, \bibinfo
  {author} {\bibfnamefont {K.}~\bibnamefont {Sharma}}, \bibinfo {author}
  {\bibfnamefont {A.}~\bibnamefont {Sornborger}}, \bibinfo {author}
  {\bibfnamefont {L.}~\bibnamefont {Cincio}}, \ and\ \bibinfo {author}
  {\bibfnamefont {P.~J.}\ \bibnamefont {Coles}},\ }\href {\doibase
  10.1038/s41467-022-32550-3} {\bibfield  {journal} {\bibinfo  {journal}
  {Nature Communications}\ }\textbf {\bibinfo {volume} {13}} (\bibinfo {year}
  {2022}),\ 10.1038/s41467-022-32550-3}\BibitemShut {NoStop}%
\bibitem [{\citenamefont {Bremner}\ \emph {et~al.}(2015)\citenamefont
  {Bremner}, \citenamefont {Montanaro},\ and\ \citenamefont
  {Shepherd}}]{Bremner2015}%
  \BibitemOpen
  \bibfield  {author} {\bibinfo {author} {\bibfnamefont {M.~J.}\ \bibnamefont
  {Bremner}}, \bibinfo {author} {\bibfnamefont {A.}~\bibnamefont {Montanaro}},
  \ and\ \bibinfo {author} {\bibfnamefont {D.~J.}\ \bibnamefont {Shepherd}},\
  }\href {\doibase 10.1103/PhysRevLett.117.080501} {\bibfield  {journal}
  {\bibinfo  {journal} {Physical Review Letters}\ }\textbf {\bibinfo {volume}
  {117}} (\bibinfo {year} {2015}),\ 10.1103/PhysRevLett.117.080501},\ \Eprint
  {http://arxiv.org/abs/1504.07999v2} {arXiv:1504.07999v2} \BibitemShut
  {NoStop}%
\bibitem [{\citenamefont {Arrasmith}\ \emph {et~al.}(2022)\citenamefont
  {Arrasmith}, \citenamefont {Holmes}, \citenamefont {Cerezo},\ and\
  \citenamefont {Coles}}]{Arrasmith2022}%
  \BibitemOpen
  \bibfield  {author} {\bibinfo {author} {\bibfnamefont {A.}~\bibnamefont
  {Arrasmith}}, \bibinfo {author} {\bibfnamefont {Z.}~\bibnamefont {Holmes}},
  \bibinfo {author} {\bibfnamefont {M.}~\bibnamefont {Cerezo}}, \ and\ \bibinfo
  {author} {\bibfnamefont {P.~J.}\ \bibnamefont {Coles}},\ }\href {\doibase
  10.1088/2058-9565/AC7D06} {\bibfield  {journal} {\bibinfo  {journal} {Quantum
  Science and Technology}\ }\textbf {\bibinfo {volume} {7}},\ \bibinfo {pages}
  {045015} (\bibinfo {year} {2022})},\ \Eprint
  {http://arxiv.org/abs/2104.05868} {arXiv:2104.05868} \BibitemShut {NoStop}%
\bibitem [{\citenamefont {Arrasmith}\ \emph {et~al.}()\citenamefont
  {Arrasmith}, \citenamefont {Cerezo}, \citenamefont {Czarnik}, \citenamefont
  {Cincio},\ and\ \citenamefont {Coles}}]{Arrasmith}%
  \BibitemOpen
  \bibfield  {author} {\bibinfo {author} {\bibfnamefont {A.}~\bibnamefont
  {Arrasmith}}, \bibinfo {author} {\bibfnamefont {M.}~\bibnamefont {Cerezo}},
  \bibinfo {author} {\bibfnamefont {P.}~\bibnamefont {Czarnik}}, \bibinfo
  {author} {\bibfnamefont {L.}~\bibnamefont {Cincio}}, \ and\ \bibinfo {author}
  {\bibfnamefont {P.~J.}\ \bibnamefont {Coles}},\ }\href {\doibase
  10.22331/q-2021-10-05-558} {\ 10.22331/q-2021-10-05-558},\ \Eprint
  {http://arxiv.org/abs/2011.12245v2} {arXiv:2011.12245v2} \BibitemShut
  {NoStop}%
\bibitem [{\citenamefont {Larocca}\ \emph {et~al.}(2022)\citenamefont
  {Larocca}, \citenamefont {Sauvage}, \citenamefont {Sbahi}, \citenamefont
  {Verdon}, \citenamefont {Coles},\ and\ \citenamefont {Cerezo}}]{Larocca2022}%
  \BibitemOpen
  \bibfield  {author} {\bibinfo {author} {\bibfnamefont {M.}~\bibnamefont
  {Larocca}}, \bibinfo {author} {\bibfnamefont {F.}~\bibnamefont {Sauvage}},
  \bibinfo {author} {\bibfnamefont {F.~M.}\ \bibnamefont {Sbahi}}, \bibinfo
  {author} {\bibfnamefont {G.}~\bibnamefont {Verdon}}, \bibinfo {author}
  {\bibfnamefont {P.~J.}\ \bibnamefont {Coles}}, \ and\ \bibinfo {author}
  {\bibfnamefont {M.}~\bibnamefont {Cerezo}},\ }\href {\doibase
  10.1103/prxquantum.3.030341} {\bibfield  {journal} {\bibinfo  {journal}
  {{PRX} Quantum}\ }\textbf {\bibinfo {volume} {3}} (\bibinfo {year} {2022}),\
  10.1103/prxquantum.3.030341}\BibitemShut {NoStop}%
\bibitem [{\citenamefont {Sauvage}\ \emph {et~al.}(2022)\citenamefont
  {Sauvage}, \citenamefont {Larocca}, \citenamefont {Coles},\ and\
  \citenamefont {Cerezo}}]{2207.14413}%
  \BibitemOpen
  \bibfield  {author} {\bibinfo {author} {\bibfnamefont {F.}~\bibnamefont
  {Sauvage}}, \bibinfo {author} {\bibfnamefont {M.}~\bibnamefont {Larocca}},
  \bibinfo {author} {\bibfnamefont {P.~J.}\ \bibnamefont {Coles}}, \ and\
  \bibinfo {author} {\bibfnamefont {M.}~\bibnamefont {Cerezo}},\ }\href@noop {}
  {\enquote {\bibinfo {title} {Building spatial symmetries into parameterized
  quantum circuits for faster training},}\ } (\bibinfo {year} {2022}),\ \Eprint
  {http://arxiv.org/abs/arXiv:2207.14413} {arXiv:2207.14413} \BibitemShut
  {NoStop}%
\bibitem [{\citenamefont {Farhi}\ and\ \citenamefont
  {Neven}(2018)}]{1802.06002}%
  \BibitemOpen
  \bibfield  {author} {\bibinfo {author} {\bibfnamefont {E.}~\bibnamefont
  {Farhi}}\ and\ \bibinfo {author} {\bibfnamefont {H.}~\bibnamefont {Neven}},\
  }\href {\doibase 10.48550/arXiv.1802.06002} {\bibfield  {journal} {\bibinfo
  {journal} {arXiv:1802.06002}\ } (\bibinfo {year} {2018}),\
  10.48550/arXiv.1802.06002}\BibitemShut {NoStop}%
\bibitem [{\citenamefont {Schuld}(2021)}]{schuld2021supervised}%
  \BibitemOpen
  \bibfield  {author} {\bibinfo {author} {\bibfnamefont {M.}~\bibnamefont
  {Schuld}},\ }\href {\doibase 10.48550/arXiv.2101.11020} {\bibfield  {journal}
  {\bibinfo  {journal} {arXiv:2101.11020}\ } (\bibinfo {year} {2021}),\
  10.48550/arXiv.2101.11020}\BibitemShut {NoStop}%
\bibitem [{\citenamefont {Sch{\"o}lkopf}\ \emph {et~al.}(2002)\citenamefont
  {Sch{\"o}lkopf}, \citenamefont {Smola}, \citenamefont {Bach} \emph
  {et~al.}}]{scholkopf2002learning}%
  \BibitemOpen
  \bibfield  {author} {\bibinfo {author} {\bibfnamefont {B.}~\bibnamefont
  {Sch{\"o}lkopf}}, \bibinfo {author} {\bibfnamefont {A.~J.}\ \bibnamefont
  {Smola}}, \bibinfo {author} {\bibfnamefont {F.}~\bibnamefont {Bach}},  \emph
  {et~al.},\ }\href {\doibase 10.7551/mitpress/4175.001.0001} {\emph {\bibinfo
  {title} {Learning with Kernels: Support Vector Machines, Regularization,
  Optimization, and Beyond}}}\ (\bibinfo  {publisher} {MIT Press},\ \bibinfo
  {year} {2002})\BibitemShut {NoStop}%
\bibitem [{\citenamefont {Seung}\ \emph {et~al.}(1992)\citenamefont {Seung},
  \citenamefont {Sompolinsky},\ and\ \citenamefont
  {Tishby}}]{sompolinsky1992examples}%
  \BibitemOpen
  \bibfield  {author} {\bibinfo {author} {\bibfnamefont {H.~S.}\ \bibnamefont
  {Seung}}, \bibinfo {author} {\bibfnamefont {H.}~\bibnamefont {Sompolinsky}},
  \ and\ \bibinfo {author} {\bibfnamefont {N.}~\bibnamefont {Tishby}},\ }\href
  {\doibase 10.1103/PhysRevA.45.6056} {\bibfield  {journal} {\bibinfo
  {journal} {Physical Review A}\ }\textbf {\bibinfo {volume} {45}},\ \bibinfo
  {pages} {6056} (\bibinfo {year} {1992})}\BibitemShut {NoStop}%
\bibitem [{\citenamefont {Dietrich}\ \emph {et~al.}(1999)\citenamefont
  {Dietrich}, \citenamefont {Opper},\ and\ \citenamefont
  {Sompolinsky}}]{sompolinsky1999statistical}%
  \BibitemOpen
  \bibfield  {author} {\bibinfo {author} {\bibfnamefont {R.}~\bibnamefont
  {Dietrich}}, \bibinfo {author} {\bibfnamefont {M.}~\bibnamefont {Opper}}, \
  and\ \bibinfo {author} {\bibfnamefont {H.}~\bibnamefont {Sompolinsky}},\
  }\href {\doibase 10.1103/PhysRevLett.82.2975} {\bibfield  {journal} {\bibinfo
   {journal} {Physical Review Letters}\ }\textbf {\bibinfo {volume} {82}},\
  \bibinfo {pages} {2975} (\bibinfo {year} {1999})}\BibitemShut {NoStop}%
\bibitem [{\citenamefont {Mezard}\ and\ \citenamefont
  {Montanari}(2009)}]{mezard2009information}%
  \BibitemOpen
  \bibfield  {author} {\bibinfo {author} {\bibfnamefont {M.}~\bibnamefont
  {Mezard}}\ and\ \bibinfo {author} {\bibfnamefont {A.}~\bibnamefont
  {Montanari}},\ }\href {\doibase 10.1093/acprof:oso/9780198570837.001.0001}
  {\emph {\bibinfo {title} {Information, Physics, and Computation}}}\ (\bibinfo
   {publisher} {Oxford University Press},\ \bibinfo {year} {2009})\BibitemShut
  {NoStop}%
\bibitem [{\citenamefont {Advani}\ \emph {et~al.}(2013)\citenamefont {Advani},
  \citenamefont {Lahiri},\ and\ \citenamefont
  {Ganguli}}]{advani2013statistical}%
  \BibitemOpen
  \bibfield  {author} {\bibinfo {author} {\bibfnamefont {M.}~\bibnamefont
  {Advani}}, \bibinfo {author} {\bibfnamefont {S.}~\bibnamefont {Lahiri}}, \
  and\ \bibinfo {author} {\bibfnamefont {S.}~\bibnamefont {Ganguli}},\ }\href
  {\doibase 10.1088/1742-5468/2013/03/P03014} {\bibfield  {journal} {\bibinfo
  {journal} {Journal of Statistical Mechanics: Theory and Experiment}\ }\textbf
  {\bibinfo {volume} {2013}},\ \bibinfo {pages} {P03014} (\bibinfo {year}
  {2013})}\BibitemShut {NoStop}%
\bibitem [{\citenamefont {Vapnik}(1996)}]{vapnik1995nature}%
  \BibitemOpen
  \bibfield  {author} {\bibinfo {author} {\bibfnamefont {V.~N.}\ \bibnamefont
  {Vapnik}},\ }\href {\doibase 10.1007/978-1-4757-3264-1} {\emph {\bibinfo
  {title} {The Nature of Statistical Learning}}}\ (\bibinfo  {publisher}
  {Springer},\ \bibinfo {year} {1996})\BibitemShut {NoStop}%
\end{thebibliography}%

\section*{Disclaimer}

This paper was prepared for information purposes with contributions from the Global Technology Applied Research center of JPMorgan Chase. This paper is not a product of the Research Department of JPMorgan Chase or its affiliates. Neither JPMorgan Chase nor any of its affiliates make any explicit or implied representation or warranty and none of them accept any liability in connection with this paper, including, but not limited to, the completeness, accuracy, reliability of information contained herein and the potential legal, compliance, tax or accounting effects thereof. This document is not intended as investment research or investment advice, or a recommendation, offer or solicitation for the purchase or sale of any security, financial instrument, financial product or service, or to be used in any way for evaluating the merits of participating in any transaction.

The U.S.\ Government retains for itself, and others acting on its behalf, a paid-up nonexclusive, irrevocable worldwide license in this paper to reproduce, prepare derivative works, distribute copies to the public, and perform publicly and display publicly, by or on behalf of the Government. The Department of Energy will provide public access to these results of federally sponsored research in accordance with the DOE Public Access Plan.

\end{document}